\documentclass[12pt,aps,tightenlines,showpacs,amsmath,amssymb]{revtex4}

\pdfoutput=1

\usepackage{graphicx}
\usepackage{subfigure}
\graphicspath{{figs/}{figs_pdf/}}
\usepackage{bm}
\usepackage{color}

\newcommand{\n}{n} 
\newcommand{\D}{D}
\newcommand{\Time}{T}
\newcommand{\Deform}{W}

\newcommand{\op}{\hat{Q}}

\begin{document}

\title{Bounds on the mixing enhancement for a stirred binary fluid}

\author{Lennon \'O N\'araigh}
\author{Jean-Luc Thiffeault}
\email{jeanluc@mailaps.org}
\thanks{Present address: Department of Mathematics, University of
  Wisconsin, Madison, WI 53706, USA.}
\affiliation{Department of Mathematics, Imperial College London, SW7 2AZ,
United Kingdom}

\date{September 11, 2007}

\begin{abstract}
  The Cahn--Hilliard equation describes phase separation in binary
  liquids.  Here we study this equation with spatially-varying sources
  and stirring, or advection.  We specialize to symmetric mixtures and
  time-independent sources and discuss stirring strategies that
  homogenize the binary fluid.  By measuring fluctuations of the
  composition away from its mean value, we quantify the amount of
  homogenization achievable.  We find upper and lower bounds on our
  measure of homogenization using only the Cahn--Hilliard equation and
  the incompressibility of the advecting flow.  We compare these
  theoretical bounds with numerical simulations for two model flows:
  the constant flow, and the random-phase sine flow.  Using the sine
  flow as an example, we show how our bounds on composition
  fluctuations provide a measure of the effectiveness of a given
  stirring protocol in homogenizing a phase-separating binary fluid.
\end{abstract}

\maketitle

\section{Introduction}
\label{sec:intro}

Phase separation and its control have received intense interest, both
because of industrial
applications~\cite{Karim2002,Aarts2005,WangH2000} and the mathematics
involved, in particular the Cahn--Hilliard equation.  This equation
was introduced by Cahn and Hilliard~\cite{CH_orig} to describe phase
separation in a binary alloy.  Since their argument relies on the
thermodynamic free energy of mixing, the equation
is completely
general, and describes any two-component system where the mixed state
is energetically unfavorable, and where the total amount of matter is
conserved.  Thus it is used in polymer physics~\cite{Aarts2005}, in
interfacial flows~\cite{LowenTrus}, and in mathematical
biology~\cite{Murray1981}.  In this paper we discuss binary liquids,
and the order-parameter equation obtained by Cahn and Hilliard
describes the composition of a binary fluid.  A composition
$c\left(\bm{x},t\right)=0$ indicates a locally well-mixed state, while
a composition $c\left(\bm{x},t\right)\neq0$ indicates a local
abundance of one binary fluid component over another.  In the context
of binary fluids, it is important to study the influence of stirring
on the two fluid components, and we therefore introduce an advection
term into the equation.

The time evolution of the non-advected equation has been studied
extensively.  In particular, there is a proof concerning the existence
and uniqueness of solutions~\cite{Elliott_Zheng}.  Given an initial
state comprising a small perturbation around the unstable, well-mixed
state $c=0$, the system forms domains of unmixed fluid that expand or
coarsen in time as $t^{1/3}$, the Lifshitz--Slyozov law~\cite{LS,
  Zhu_numerics}.  In many applications, the coarsening tendency of the
Cahn--Hilliard fluid is undesirable, and this can be overcome by fluid
advection, which is either passive or active.  In the active
case~\cite{Bray_advphys, Berti2005, ONaraigh2007_2}, the composition
gradients induce a backreaction on the flow, while in the passive
case, the composition and its gradients exert no effect on the
flow~\cite{chaos_Berthier, shear_Berthier, ONaraigh2007, Sancho}, and
it is this case that we consider in the present paper.  If the flow
contains large differential shears, the coarsening is either arrested,
so that domains grow only to a certain size~\cite{chaos_Berthier}, or
can be overwhelmed entirely, so that the mixed state, previously
unstable, is now favored~\cite{ONaraigh2007}.  This homogenization is
useful in applications, for example in the fabrication of emulsion
paints~\cite{Karim2002}.

Stirring provides one means of controlling phase separation.  Indeed,
it is well known that a unidirectional shear flow produces banded
domains, with domains aligned in the shear
direction~\cite{shear_Berthier, shear_Bray, shear_Shou, Hashimoto}.
Other mechanisms have been employed to control the phase separation,
including dipole interactions~\cite{Lu2005}, patterned
substrates~\cite{Lu2000}, and surface forcing~\cite{Kielhorn1999}.
In~\cite{ONaraigh2007_2}, \'O N\'araigh and Thiffeault study binary
mixtures in thin films, and use the backreaction, together with
surface forcing, to align the domains in a direction that depends on
the forcing parameters.  In the present paper, we focus on phase
separation in the presence of advection and a spatially-varying
persistent source.  This source can be maintained in several ways.
In~\cite{Krekhov2004} the spatial source is produced by thermal
diffusion, through the Ludwig--Soret effect~\cite{Craig2004}, in which
composition gradients are induced by imposed temperature gradients.
One can also produce composition gradients simply by injecting matter
into the system~\cite{lattice_PH2}.  In this paper, we shall restrict
our attention to the symmetric binary fluid, in which equal amounts of
both fluid components are present.  This special case involves a wide
range of
phenomena~\cite{Berti2005,Krekhov2004,Zhu_numerics,Hashimoto,Tong1989}.
In the present work, we shall refer to the combination of stirring
(advection) and the spatial source as `forcing.'

We have mentioned the often undesirable coarsening tendency of the binary
fluid, and the efforts made to suppress it.  
In this paper we introduce a quantitative measure of coarsening suppression
by studying the $p^{\mathrm{th}}$ power-mean fluctuation of the composition
about its average value.  
By fluctuations about the average value, we mean spatial fluctuations around
the (constant) average spatial composition, which we then average over space
and time.
In effect, we study the time-averaged $L^p$ norm of the composition.
If this quantity is small, the average deviation of the system about the
well-mixed state $c=0$ is small, and we therefore use this quantity as a
proxy for the level of mixedness of the fluid.
An approach similar to this has already been taken for miscible
fluids~\cite{DoeringThiffeault2006,Shaw2007,Thiffeault2004,Thiffeault2006,Thiffeault2007}.
There, the equation of interest is the advection-diffusion equation,
and fluctuations about the mean are measured by the variance or
centred second power-mean of the fluid
concentration~\cite{Danckwerts1952,Edwards1985}.
The variance is reduced by stirring.  By specifying a
source term, it is possible to state the maximum amount by which a given
flow can reduce the variance, and hence mix the fluid.  
By quantifying the variance reduction, one can classify flows according to
how effective they are at mixing.
Just as the linearity of the advection-diffusion equation suggests the
variance as a natural way of measuring fluctuations in the
concentration, the non-linearity of the Cahn--Hilliard equation and
its associated free energy (cubic and quartic in the composition,
respectively), will fix our attention on the fourth power-mean of the
composition fluctuations.  Owing to H\"older's inequality, a binary
liquid that is well-mixed in this sense will also be well-mixed in the
variance sense.  The advantage we gain in considering the fourth
power-mean is the derivation of explicit bounds on our measure of
mixedness which have manifest flow- and source-dependence.

The paper is organized as follows.  In Sec.~\ref{sec:model} we
introduce the model equations and discuss their nondimensionalization.
We introduce measures of composition fluctuations and their relation
to fluid mixing.  These measures of composition fluctuations are based
on long-time averages obtained from the composition of the binary
fluid.  Therefore, in Sec.~\ref{sec:existence} we prove the existence
of these long-time averages and find upper bounds on the measures of
composition fluctuations.  Since we are interested in minimizing
composition fluctuations, in Sec.~\ref{sec:lower_bds} we obtain lower
bounds on these measures.  In Sec.~\ref{sec:scaling} we investigate
the parametric dependence of the upper and lower bounds for
statistical, homogeneous, isotropic turbulence.  In
Sec.~\ref{sec:numerics} we compare the theoretical bounds with
numerical simulations for two standard flows: the constant flow, and
the sine flow.  In the numerical simulations, we find that the
composition fluctuations are indeed bounded by the theoretical limits
we have obtained, and the results are dramatically different from
those obtained for miscible liquids.


\section{The model equations}
\label{sec:model}

In this section, we introduce the advective Cahn--Hilliard equation
with sources and discuss its properties.  We outline the tools and
notation we shall use to analyze composition fluctuations.  For
generality, the discussion takes place against the backdrop of scalar
and vector fields in $\mathbb{R}^n$.

Let $c\left(\bm{x}, t\right)$ be the composition, that is, the scalar
field $c\left(\bm{x},t\right)$ measures phase separation, with
$c\left(\bm{x},t\right)=0$ indicating a locally well-mixed state, and
$c\left(\bm{x},t\right)\neq0$ indicating a local excess of one binary
fluid component relative to the other.  Let
$\bm{v}\left(\bm{x},t\right)$ be an externally imposed $n$-dimensional
incompressible flow, $\nabla\cdot\bm{v}=0$, and let
$s\left(\bm{x}\right)$ be a distribution of sources and sinks of
binary fluid.  The advective Cahn--Hilliard (ACH) equation describes
the phase-separation dynamics of the scalar field $c\left(\bm{x},
t\right)$ in the presence of flow, for prescribed sources and sinks, is
\begin{equation}
\frac{\partial c}{\partial t}+\bm{v}\cdot\nabla c = D\Delta\left(c^3-c-\gamma\Delta
c\right)+s\left(\bm{x}\right).\\
\label{eq:ch}
\end{equation}
Here $D$ is Cahn--Hilliard diffusion
coefficient and $\sqrt{\gamma}$ is the typical thickness of transition zones
between phase-separated regions of the binary fluid.  The finite thickness
of these zones prevents the formation of infinite gradients in the problem.
The equation is a passive advection equation: neither the composition
nor its gradients affect the flow.

In this paper we work with a nondimensionalization of
Eq.~\eqref{eq:ch} that leaves three control parameters in the problem.
Therefore, we can unambiguously study limits where control parameters
tend to zero or infinity.  Let $\Time$ be a timescale associated with
the velocity $\bm{v}\left(\bm{x},t\right)$, and let $V_0$ be the
magnitude of $\bm{v}\left(\bm{x},t\right)$.  Let $S_0$ be the
magnitude of the source variations and, finally, let $L$ be a
lengthscale in the problem; for example, if the problem is solved in a
cube with periodic boundary conditions, we take the lengthscale $L$ to
be the cube length.  It is then possible to write down
Eq.~\eqref{eq:ch} using a nondimensional time $t'=t/\Time$ and a
nondimensional spatial variable $\bm{x}'=\bm{x}/L$,
\begin{equation*}
\frac{\partial c}{\partial t'}+V_0'\tilde{\bm{v}}\cdot\nabla' c = D'\Delta'\left(c^3-c-\gamma'\Delta'
c\right)+S_0'\tilde{s}\left(\bm{x}'\right),\\
\end{equation*}
$\tilde{\bm{v}}$ and $\tilde{s}$ are dimensionless shape functions,
$V_0' = \Time V_0/L$, $D' = D\Time/L^2$, $\gamma'=\gamma/L^2$, and
where $S_0'=S_0\Time$.  The quantity $D' = D\Time/L^2$ is identified
with~$\Time/\Time_D$, the ratio of the velocity timescale to the
diffusion timescale.  Following standard practice, we shall now work
with the dimensionless version of the equation, and omit the prime
notation.  For ease of notation, we shall henceforth take $\bm{v}$ to
mean $V_0'\tilde{\bm{v}}$.

 The equation~\eqref{eq:ch} has the following properties, which
 we shall exploit in our analysis:
\begin{itemize}
\item If the source $s\left(\bm{x}\right)$ is chosen to have spatial mean
zero, then the total mass is conserved,
\[
\frac{d}{dt}\int_{\Omega}c\left(\bm{x},t\right)d^n x = \int_{\Omega}D\Delta\left(c^3-c-\gamma\Delta
c\right)d^n x+\int_{\Omega}s\left(\bm{x}\right)d^n x=0+\left[\text{boundary
terms}\right],
\]
where $\Omega$ is the problem domain in $n$ dimensions and $|\Omega|$
is its volume.  The boundary terms in this equation vanish on choosing
no-flux boundary conditions
$\hat{\bm{n}}\cdot\nabla{c}=\hat{\bm{n}}\cdot\nabla\mu=0$ on
$\partial\Omega$, or periodic boundary conditions.  Here
$\hat{\bm{n}}\cdot\nabla$ is the outward normal derivative.  In this
paper we shall consider the periodic case.
\item There is a free-energy functional
\begin{equation}
F\left[c\right]=\int_\Omega\left[\tfrac{1}{4}\left(c^2-1\right)^2+\tfrac{1}{2}\gamma\left|\nabla
c\right|^2\right]d^nx,\qquad\mu=\frac{\delta F}{\delta c}=c^3-c-\gamma\Delta
c,
\label{eq:fe}
\end{equation}
where $\mu$ is the chemical potential of the system~\cite{Cahn1965}.
For a smooth composition $c\left(\bm{x},t\right)$, the free energy satisfies
the evolution equation
\[
\dot{F}\equiv\frac{dF}{dt}=-D\int_\Omega\left|\nabla\mu\right|^2d^nx+\int_\Omega\mu\left(-\bm{v}\cdot\nabla
c+s\right)d^{n}x,
\]
and decays in time in the absence of sources and stirring.
\end{itemize}

To study the spatial fluctuations in composition, we consider the
power-means of the quantity
$c\left(\bm{x},t\right)-|\Omega|^{-1}\int_\Omega{c\left(\bm{x},t\right)}d^nx$,
\begin{equation}
{M}_p\left(t\right)=\bigg\{\int_\Omega\left|c\left(\bm{x},t\right)-\frac{1}{|\Omega|}\int_{\Omega}c\left(\bm{x},t\right)d^nx\right|^p\bigg\}^{\frac{1}{p}}.
\label{eq:power_mean}
\end{equation}  
For a symmetric mixture in which $\int_{\Omega}c\left(\bm{x},t\right)d^{\n}x=0$,
this is simply
\[
{M}_p\left(t\right)=\bigg\{\int_\Omega\left|c\left(\bm{x},t\right)\right|^pd^nx\bigg\}^{\frac{1}{p}}=\|c\|_p,
\]
where we have introduced the $L^p$ norm of the composition, $\|c\|_p$.  The
quantity $M_p$ is a measure of the magnitude of spatial fluctuations in the
composition about the mean, at a given time.  Since
we are interested in the ultimate state of the system, we study the long-time
average of composition fluctuations.  We therefore focus on the power-mean
fluctuations
\[
m_p=\langle M_p^p\rangle^{\frac{1}{p}},
\]
where $\langle\cdot\rangle$ is the long-time average
\[
\langle\cdot\rangle=\lim_{t\rightarrow\infty}\frac{1}{t}\int_0^t\left(\cdot\right){ds},
\]
provided the limit exists.  We shall repeatedly use the following results
for the monotonicity of norms,
\begin{eqnarray}
\|f\|_p&\leq&\left|\Omega\right|^{\frac{1}{p}-\frac{1}{q}}\|f\|_q,\qquad\qquad
1\leq p\leq q,\ f\in L^q(\Omega),\nonumber\\
\frac{1}{t}\int_0^t \left|g\left(s\right)\right|ds&\leq&
\left[\frac{1}{t}\int_0^t\left|g\left(s\right)\right|^q ds\right]^{\frac{1}{q}},
\qquad q\geq 1,\ g\in L^s\left(\left[0,t\right]\right),
\label{eq:monotonicity_norms}
\end{eqnarray}
which follow from the H\"older inequality.

The Cahn--Hilliard equation and its free energy functional contain high powers
of the composition $c$ ($c^3$ and $c^4$ respectively), and we can therefore
estimate $m_p$ for specific $p$-values.  In particular, in the following
sections, we shall prove the following result in $\n$ dimensions:
\begin{quote}
  Given a smooth solution to the ACH equation, the long-time average
  of the free energy exists, and therefore $m_p$ exists, for
  $1 \le p \le 4$.
\end{quote}
The constraints we impose on the forcing terms are that the velocity
field and its first spatial derivatives be bounded in the $L^{\infty}$
norm, and that the source term be bounded in the $L^2$ norm.  That is,
$\bm{v}\in L^{\infty}\left(0,T;H^{1,\infty}\left(\Omega\right)\right)$
for any $T\in\left[0,\infty\right)$, and $s\in
L^2\left(\Omega\right)$.
%
%
%
%
%
We take our
result one step further by explicitly evaluating upper and lower bounds for
$m_4$, and this gives a way of quantifying composition fluctuations in the
stirred binary fluid.

\section{Existence of long-time averages}
\label{sec:existence}

In this section, we prove a result concerning the
existence of the long-time average of the free energy, and of the power-means
$m_p$, for $1 \le p \le 4$.
\begin{quote}
  Given the velocity field $\bm{v}\left(\bm{x},t\right)\in L^{\infty}\left(0,T;H^{1,\infty}\left(\Omega\right)\right)$
  for any $T\in\left[0,\infty\right)$, the source $s\left(\bm{x}\right)\in
  L^2\left(\Omega\right)$, and smooth initial data for the ACH equation~\eqref{eq:ch},
  the long-time average of the free energy exists, and thus $m_p$ exists,
  for $1 \le p \le 4$.
\end{quote}
The proof relies on the free-energy evolution equation.  Using this
law, we find uniform bounds on the finite-time means $\langle
F\rangle_t$ and $\langle M_4^4\rangle_t$, where
\[
\langle\cdot\rangle_t=\frac{1}{t}\int_0^t\left(\cdot\right)ds,\qquad\langle\cdot\rangle=\lim_{t\rightarrow\infty}\langle\cdot\rangle_t.
\]
Using the monotonicity of norms, the uniform boundedness of $\langle
M_p^p\rangle_t$, follows, for $1\leq p\leq4$.  The proof proceeds in
multiple steps, which we outline below.

\subsection*{Step 1: Analysis of the free-energy evolution equation}

By modifying the argument of Elliott and Zheng~\cite{Elliott_Zheng}
for the Cahn--Hilliard equation without flow and sources, it is
readily seen that for smooth initial data, and for forcing terms with
the regularity properties just mentioned, a unique smooth solution to
the ACH equation exists, at least for finite times.  Thus, we turn to
the question of the long-time behaviour of solutions.
We exploit the smoothness properties of the composition field $c\left(\bm{x},t\right)$
and formulate an evolution equation for the free energy
\[
F\left[c\right]=\int_\Omega\left[\tfrac{1}{4}\left(c^2-1\right)+\tfrac{1}{2}\gamma\left|\nabla{c}\right|^2\right]d^{\n}x.
\]
Given the smooth, finite-time solution $c\left(\bm{x},t\right)$, we
differentiate the functional $F\left[c\right]$ with respect to time and obtain
the relation
\[
\frac{dF}{dt}=\int_{\Omega}\frac{\partial c}{\partial t}\mu\,
d^{\n}x,\qquad\mu=c^3-c-\gamma\Delta{c}.
\]
Since $c\left(\bm{x},t\right)$ satisfies the ACH equation~\eqref{eq:ch},
the evolution equation takes the form
\[
\frac{dF}{dt}=-D\int_\Omega\left|\nabla\mu\right|^2d^nx+\int_\Omega\mu\left(-\bm{v}\cdot\nabla
c+s\right)d^{n}x,
\]
using the no-flux or periodic boundary conditions.  By averaging
this equation over finite times, we obtain the identity
\begin{equation}
\langle{\dot{F}}\rangle_t+\D\bigg\langle\int_\Omega\left|\nabla\mu\right|^2d^\n
x\bigg\rangle_t=\bigg\langle\int_\Omega\mu
sd^\n x\bigg\rangle_t-\bigg\langle\int_\Omega\mu\bm{v}\cdot\nabla cd^\n x\bigg\rangle_t,
\label{eq:fe_dissipation}
\end{equation}
We single out the quantity $\langle\dot{F}\rangle$ for study.  Owing
to the nonnegativity of $F\left(t\right)$, we have the inequality
$\langle\dot{F}\rangle\geq0$. Therefore, we need only consider two
separate cases: $\langle\dot{F}\rangle=0$, and
$\langle\dot{F}\rangle>0$.  We shall show that
$\langle\dot{F}\rangle\neq0$ is not possible, and in doing so, we
shall produce a uniform ($t$-independent) upper bound on
$\langle{F}\rangle_t$.

Let us assume for contradiction that $\langle\dot{F}\rangle>0$.  Then,
given any $\varepsilon$ in the range
$0<\varepsilon<\langle\dot{F}\rangle$, there is a time $T_\varepsilon$
such that
$\langle\dot{F}\rangle-\varepsilon<\langle{\dot{F}}\rangle_t<\langle\dot{F}\rangle+\varepsilon$,
for all times $t>T_\varepsilon$.  Thus, for times $t>T_\varepsilon$,
the time average $\langle{\dot{F}}\rangle_t$ is strictly positive.
Henceforth, the inequality $t>T_\varepsilon$ is assumed.  We use the condition
$\nabla\cdot\bm{v}=0$, together with integration by parts, and find the last
term in Eq.~\eqref{eq:fe_dissipation} becomes
\begin{equation}
\langle{\dot{F}}\rangle_t+\D\bigg\langle\int_\Omega\left|\nabla\mu\right|^2d^\n
x\bigg\rangle_t=\bigg\langle\int_\Omega\mu\, s\,d^\n x\bigg\rangle_t+\bigg\langle\gamma\int_\Omega\Delta
c\bm{v}\cdot\nabla cd^\n x\bigg\rangle_t.
\label{eq:dissipation}
\end{equation}
Now
\begin{align*}
\int_\Omega\Delta c\,\bm{v}\cdot\nabla cd^\n x
&=
-\int_\Omega\left(\partial_ic\right)\left[\partial_i\left(v_j\partial_jc\right)\right]d^{\n}x,\\
&=-\int_\Omega\left(\partial_ic\right)\left(\partial_iv_j\right)\left(\partial_jc\right)d^{\n}x-
\int_\Omega\left(\partial_ic\right)\left(\bm{v}\cdot\nabla\right)\left(\partial_ic\right)d^{\n}x,\\
&=-\int_\Omega\bm{w}\Deform\bm{w}^Td^\n x,
\end{align*}
with
\begin{equation*}
\qquad\bm{w}=\nabla c,\qquad
{\Deform}_{ij}=\tfrac{1}{2}\left(\partial_iv_j+\partial_jv_i\right),
\end{equation*}
where we have used the summation convention for repeated indices
and have omitted terms in the integration identities that vanish as a result
of our choice of boundary conditions.
The quadratic form $\bm{w}\Deform\bm{w}^T$ satisfies 
$\left|\bm{w}\Deform\bm{w}^T\right|\leq{\n}\max_{ij}|\Deform_{ij}|\|\bm{w}\|_2^2$,
which gives rise to the inequality
\begin{equation}
\left|\int_\Omega\Delta c\bm{v}\cdot\nabla cd^\n x\right|\leq{\n}\left(\sup_{\Omega,i,j}\left|\Deform_{ij}\right|\right)\int\left|\nabla
c\right|^2d^\n x.
\label{eq:bound_on_bad_bit}
\end{equation}
%
%
%
%
%
%
The matrix ${\Deform}$ is the rate-of-strain tensor.  The appearance
of the rate-of-strain tensor in our analysis shows the importance of
shear and stretching in the development of the composition morphology.

For each time $t'\in\left[0,t\right]$, we split the chemical potential
$\mu$ into a part with mean zero, and a mean component: $\mu =
\overline{\mu}\left(t'\right)+\mu'\left(\bm{x},t'\right)$, where
$\int_\Omega\mu'\left(\bm{x},t'\right)d^n x=0$.  Then, for any
function $\phi\left(\bm{x},t\right)$ with spatial mean zero, we have
the relation $\int_\Omega\phi\mu\, d^{\n}x=\int_\Omega\phi\mu'd^{\n}x$.
Using this device, Eq.~\eqref{eq:dissipation} becomes
\[
\langle{\dot{F}}\rangle_t+\D\bigg\langle\int_\Omega\left|\nabla\mu'\right|^2d^\n
x\bigg\rangle_t=
\bigg\langle\int_\Omega\mu' sd^\n x\bigg\rangle_t+\bigg\langle\gamma\int_\Omega\Delta
c\bm{v}\cdot\nabla cd^\n x\bigg\rangle_t.
\]
Owing to the positivity of $\langle{\dot{F}}\rangle_t$, we have the inequality
\begin{equation}
\D\bigg\langle\int_\Omega\left|\nabla\mu'\right|^2d^{\n}x\bigg\rangle_t\leq
%
%
\bigg\langle\int_\Omega\mu' sd^\n x\bigg\rangle_t+\bigg\langle\gamma\int_\Omega\Delta
c\bm{v}\cdot\nabla cd^\n x\bigg\rangle_t.
\label{eq:dissipation1}
\end{equation}
Finally, we employ the Poincar\'e inequality for mean-zero functions
on a periodic domain $\Omega=\left[0,L\right]^2$,
%
%
\begin{equation}
\|\mu'\|_2^2\leq \left(\frac{L}{2\pi}\right)^2\|\nabla \mu'\|_2^2.
\label{eq:Poincare}
\end{equation}
Combining Eqs.~\eqref{eq:bound_on_bad_bit},~\eqref{eq:dissipation1}, and~\eqref{eq:Poincare}
gives the following inequality:
\begin{equation}
\D\left(\frac{2\pi}{L}\right)^2\langle\|\mu'\|_2^2\rangle_t\leq
\langle\|\mu'\|_2^2\rangle_t^{\frac{1}{2}}\|s\|_2+
\n \Deform_{\infty}\bigg\langle\int_\Omega\gamma\left|\nabla c\right|^2
d^\n x\bigg\rangle_t,
\label{eq:inequality_for_mu}
\end{equation}
where $\Deform_{\infty}=\text{sup}_{t,\Omega,i,j}\left|\Deform_{ij}\right|$.
 There are no angle brackets around the source term because $s\left(\bm{x}\right)$
 is independent of time.

\subsection*{Step 2: Obtaining a bound on $\langle\|\mu'\|_2^2\rangle_t$}

Using
\[
\int_\Omega\gamma\left|\nabla c\right|^2d^{\n}x
%
%
=\int_\Omega\left[\mu' c+c^2-c^4\right]d^{\n}x
%
%
\leq\|\mu'\|_2\|c\|_2+\|c\|_2^2,
\]
we obtain the inequality 
\begin{equation}
\int\gamma\left|\nabla c\right|^2d^\n x\leq\left|\Omega\right|^{\frac{1}{4}}\|\mu'\|_2\|c\|_4+\left|\Omega\right|^{\frac{1}{2}}\|c\|_4^2.
\label{eq:inequality_grad_c}
\end{equation}
Combining Eqs.~\eqref{eq:inequality_for_mu} and~\eqref{eq:inequality_grad_c},
\[
\D\left(\frac{2\pi}{L}\right)^2\langle\|\mu'\|_2^2\rangle_t\leq
\langle\|\mu'\|_2^2\rangle_t^{\frac{1}{2}}\left[\|s\|_2+\n\left|\Omega\right|^{\frac{1}{4}}\Deform_\infty\langle\|c\|_4^4\rangle_t^{\frac{1}{4}}\right]+
2\left|\Omega\right|^{\frac{1}{2}}\Deform_\infty\langle\|c\|_4^4\rangle_t^{\frac{1}{2}},
\]
a quadratic inequality in $\langle\|\mu'\|_2\rangle_t^{\frac{1}{2}}$.  Hence,
\begin{multline*}
\langle\|\mu'\|_2^2\rangle_t^{\frac{1}{2}}\leq\frac{1}{2D}\left(\frac{L}{2\pi}\right)^2
\left(\|s\|_2+\n\left|\Omega\right|^{\frac{1}{4}}\Deform_{\infty}\langle\|c\|_4^4\rangle_t^{\frac{1}{4}}\right)\\
+
\frac{1}{2D}\left(\frac{L}{2\pi}\right)^2
\left[\left(\|s\|_2+\n\left|\Omega\right|^{\frac{1}{4}}\Deform_{\infty}\langle\|c\|_4^4\rangle_t^{\frac{1}{4}}\right)^2+{8D\left|\Omega\right|^{\frac{1}{2}}}\left(\frac{2\pi}{L}\right)^2\Deform_\infty\langle\|c\|_4^4\rangle_t^{\frac{1}{2}}\right]^{\frac{1}{2}}.
\end{multline*}
A less sharp bound is given by
%
%
%
%
%
%
%
%
%
%
%
\begin{equation}
\langle\|\mu'\|_2^2\rangle_t\leq
\frac{1}{D^2}\left(\frac{L}{2\pi}\right)^4 
\left(\|s\|_2+\n\left|\Omega\right|^{\frac{1}{4}}\Deform_\infty\langle\|c\|_4^4\rangle_t^{\frac{1}{4}}\right)^2+\frac{8\left|\Omega\right|^{\frac{1}{2}}}{D}\left(\frac{L}{2\pi}\right)^2\Deform_\infty\langle\|c\|_4^4\rangle_t^{\frac{1}{2}},
\label{eq:bound_on_mu}
\end{equation}
which is an upper bound for $\langle\|\mu'\|_2^2\rangle_t$, in terms of the
forcing parameters and $\langle\|c\|_4^4\rangle_t$.
\subsection*{Step 3: An upper bound on $m_4$}
We have the free energy
\begin{multline*}
F\left[c\right]=\int_\Omega\left[\tfrac{1}{4}\left(c^2-1\right)^2+\tfrac{1}{2}\gamma\left|\nabla
c\right|^2\right]d^\n x = 
\int_\Omega\left[\tfrac{1}{2}c\mu-\tfrac{1}{4}c^4\right]d^\n x+\tfrac{1}{4}\left|\Omega\right|\\
%
=\int_\Omega\left[\tfrac{1}{2}c\mu'\left(\bm{x},t'\right)-\tfrac{1}{4}c^4\right]d^{\n}x+\tfrac{1}{4}\left|\Omega\right|\geq0.
\end{multline*}
Hence,
\[
\int_\Omega c^4d^\n x\leq2\int_\Omega c\,\mu' d^\n x + \left|\Omega\right|
\leq2\|c\|_2\|\mu'\|_2+\left|\Omega\right|.
\]
Time-averaging both sides and using the monotonicity of
norms~\eqref{eq:monotonicity_norms}, we obtain the result
\[
\langle\|c\|_4^4\rangle_t\leq\left|\Omega\right|+2\left|\Omega\right|^{\frac{1}{4}}\langle\|c\|_4^4\rangle_t^{\frac{1}{4}}\langle\|\mu'\|_2^2\rangle_t^{\frac{1}{2}}.
\]
Using the bound for $\langle\|\mu'\|_2^2\rangle_t$ in~\eqref{eq:bound_on_mu},
this becomes
\begin{multline*}
\langle\|c\|_4^4\rangle_t\leq\left|\Omega\right|
\\
+\frac{2\left|\Omega\right|^{\frac{1}{4}}}{D}\left(\frac{ L}{2\pi}\right)^2\langle\|c\|_4^4\rangle_t^{\frac{1}{4}}\left[\left(\|s\|_2+\n\left|\Omega\right|^{\frac{1}{4}}\Deform_\infty
\langle\|c\|_4^4\rangle_t^{\frac{1}{4}}\right)^2+{4\n D\left|\Omega\right|^{\frac{1}{2}}}\left(\frac{2\pi}{L}\right)^2\Deform_\infty\langle\|c\|_4^4\rangle_t^{\frac{1}{2}}\right]^{\frac{1}{2}}.
\end{multline*}
We therefore have a $t$-independent equation for the upper bound on $\langle\|c\|_4^4\rangle_t$,
\[
\langle\|c\|_4^4\rangle_t\leq m_{4}^{\mathrm{max}}\left(\bm{v},s,D\right),
\]
where $m_4^{\mathrm{max}}$ solves the polynomial
\begin{multline}
\left(m_{4}^{\mathrm{max}}\right)^4=\left|\Omega\right|
\\
+\frac{2\left|\Omega\right|^{\frac{1}{4}}}{D}\left(\frac{L}{2\pi}\right)^2m_{4}^{\mathrm{max}}\left[\left(\|s\|_2+\n\left|\Omega\right|^{\frac{1}{4}}\Deform_\infty{m_{4}}^{\mathrm{max}}\right)^2+{4\n
D\left|\Omega\right|^{\frac{1}{2}}}\left(\frac{2\pi}{L}\right)^2\Deform_\infty\left({m_{4}}^{\mathrm{max}}\right)^2\right]^{\frac{1}{2}},
\label{eq:eqn_for_x}
\end{multline}
The highest power of $m_{4}^{\mathrm{max}}$ on the left-hand side is
$\left(m_{4}^{\mathrm{max}}\right)^4$, while the highest power of
$m_{4}^{\mathrm{max}}$ on the right-hand side is
$\left(m_{4,t}^{\mathrm{max}}\right)^{\frac{3}{2}}$.  Thus, this
equation always has a unique positive solution.
 
We obtain the following chain of uniform ($t$-independent) bounds.  Each
bound follows from the previous bounds in the chain, and the first bound
follows from Eq.~\eqref{eq:eqn_for_x}.
\begin{itemize}
\item $\langle\|c\|_4^4\rangle_t$ is uniformly bounded,
\item $\langle\|c\|_2^2\rangle_t$ is uniformly bounded,
\item $\langle\|\mu'\|_2^2\rangle_t$ is uniformly bounded,\vskip-0.45in\begin{equation}\label{eq:bounds}\end{equation}
\item $\langle\|\nabla{c}\|_2^2\rangle_t$ is uniformly bounded,
\item $\langle F\rangle_t$ is uniformly bounded,
\end{itemize}
for all $t>T_\varepsilon$.  
Owing to the uniformity of these bounds, they hold in the limit
$t\rightarrow\infty$.  The result $\langle F\rangle<\infty$ implies
the existence of a uniform bound for $F\left(t\right)$, almost
everywhere.  Given the differentiability of $F\left(t\right)$, this
implies that $F\left(t\right)$ is everywhere uniformly bounded, and
thus, $\langle\dot{F}\rangle=0$, which is a contradiction.  Therefore,
the only possibility for $\langle\dot{F}\rangle$ is that it be zero.
It is straightforward to verify that by taking
$\langle\dot{F}\rangle=0$, and making slight alterations in steps
1--3, the bounds in Eq.~\eqref{eq:bounds} still hold.

Let us examine the significance of our result.  We have shown that for
sufficiently regular flows and source terms (specifically,
$\bm{v}\left(\bm{x},t\right)\in
L^{\infty}\left(0,T;H^{1,\infty}\left(\Omega\right)\right)$,
$T\in\left[0,\infty\right)$, and $s\left(\bm{x}\right)\in
L^2\left(\Omega\right)$), there is an \emph{a priori} bound on the
compositional free energy $\langle F\left[c\right]\rangle$.  We have
shown that the system always reaches a steady state, in the sense that
$\langle\dot{F}\rangle=0$.  We have also found an upper bound for the
$m_4$ measure of composition fluctuations, as the unique positive root
of the polynomial equation Eq.~\eqref{eq:eqn_for_x}.  This bound
depends only on the source amplitude, the diffusion constant, and the
maximum rate-of-strain $\Deform_\infty$.  Using the monotonicity of
norms, this number serves also as an upper bound on $m_p$ for $1 \le p
\le 4$.  Let us comment briefly on the volume term in the equation
$\left(m_{4}^{\mathrm{max}}\right)^4=\left|\Omega\right|+...$.  Since
this upper bound includes many situations, it must take into account
the case where both the velocity and the source vanish.  Then
$c\sim\pm1$ as $t\rightarrow\infty$, and by definition,
$m_4\sim|\Omega|^{\frac{1}{4}}$, which is in agreement with
Eq.~\eqref{eq:eqn_for_x}.
 
As mentioned in Sec.~\ref{sec:intro}, it is desirable in many
applications to suppress composition fluctuations, since this leads to
a homogeneous mixture.  In this paper, we propose advection as a
suppression mechanism, and we would therefore like to know the maximum
suppression achievable for a given flow.  This suggests that we seek
lower bounds on $m_p$, in addition to the upper bounds found in this
section.

\section{Lower bounds on the composition fluctuations}
\label{sec:lower_bds}

In this section we discuss the significance of the lower bound
on the measure $m_p$ of composition fluctuations.  Due to the powers of the
composition that appear in the Cahn--Hilliard equation, it is possible
to obtain an explicit lower bound for $m_4$, which we then use to discuss
mechanisms to suppress composition fluctuations.  After taking care of the
volume factors, the lower bound on $m_4$ must be greater than or equal to
the lower bound on $m_p$, for $1 \le p \le 4$.  Thus, a flow that
suppresses composition fluctuations in the $m_4$ sense will also suppress
them in in the $m_p$ sense, for $1 \le p \le 4$.

As discussed in Sec.~\ref{sec:model}, a suitable
measure of composition fluctuations for a symmetric mixture is
\[
m_p=\langle\|c\|_p^p\rangle^{\frac{1}{p}},
\]
where $c\left(\bm{x},t\right)$ is the composition of the binary
mixture and $\|c\|_p$ is its $L^p$ norm.  For $p=2$, this gives the
usual variance, used in the theory of miscible fluid
mixing~\cite{DoeringThiffeault2006,Shaw2007,Thiffeault2004,Thiffeault2006,Thiffeault2007}.
In that case, the choice $p=2$ is a natural one suggested by the
linearity of the advection-diffusion equation.  In the following
analysis of the ACH equation, it is possible to find an explicit lower
bound for $m_4$ and we therefore use this quantity to study the
suppression of composition fluctuations due to the imposed velocity
field.  Given this formula, we can compare the suppression achieved by
a given flow with the ideal level of suppression, and decide on the
best strategy to homogenize the binary fluid.

To estimate $m_4$, we take the ACH equation~\eqref{eq:ch}, multiply it
by an arbitrary, spatially-varying test function $\phi\left(\bm{x}\right)$,
and then integrate over space and time, which yields
\begin{equation}
\Big\langle\int_\Omega c\left[\op\phi+Dc^2\Delta\phi\right]d^nx\Big\rangle
=-\int_{\Omega}s{\phi}\,d^{\n}x,
\label{eq:constraint_exact}
\end{equation}
where $\op$ is the linear operator $\bm{v}\cdot\nabla - D\Delta-\gamma
D\Delta^2$.  Using the constraint Eq.~\eqref{eq:constraint_exact}, the
monotonicity of norms, and the Cauchy--Schwarz inequality, we obtain
the following string of inequalities,
\[
\left|\int_\Omega s\phi d^\n x\right|
%
%
\leq\langle\|c\|_2\|\op\phi+\D{c^2}\Delta\phi\|_2\rangle
\leq\langle\|c\|_2^2\rangle^{\frac{1}{2}}\langle\|\op\phi+\D c^2\Delta\phi\|_2^2\rangle^{\frac{1}{2}},
\]
which gives the relation
\[
\langle\|c\|_2^2\rangle^{\frac{1}{2}}\geq\frac{\left|\int_\Omega s{\phi}d^{\n}x\right|}
{\langle\|\op\phi+\D{c^2}\Delta\phi\|_2^2\rangle^{\frac{1}{2}}}.
\]
We study the denominator
\begin{eqnarray*}
\langle\|\op\phi+\D{c^2}\Delta\phi\|_2^2\rangle^{\frac{1}{2}}&\leq&\langle\|\op\phi\|_2^2\rangle^{\frac{1}{2}}+\D\langle\|c^2\Delta\phi\|_2^2\rangle^{\frac{1}{2}},\\
&\leq&\Big\langle\int_\Omega(\op\phi)^2d^\n x\Big\rangle^{\frac{1}{2}}+\D\|\Delta\phi\|_\infty\langle\|c\|_4^4\rangle^{\frac{1}{2}},
\end{eqnarray*}
where this bound follows from the triangle and H\"older inequalities.
 Thus we have the result
\[
\langle\|c\|_2^2\rangle^{\frac{1}{2}}\geq\frac{\left|\int_\Omega s\phi d^{\n}x\right|}{\langle\int_\Omega(\op\phi)^2d^\n
x\rangle^{\frac{1}{2}}+\D\|\Delta\phi\|_\infty\langle\|c\|_4^4\rangle^{\frac{1}{2}}},
\]
or
\[
\langle\|c\|_2^2\rangle^{\frac{1}{2}}\left[\Big\langle\int_\Omega(\op\phi)^2d^{\n}x\Big\rangle^{\frac{1}{2}}
+\D\|\Delta\phi\|_\infty\langle\|c\|_4^4\rangle^{\frac{1}{2}}\right]\geq\left|\int_\Omega
s\phi d^\n x\right|.
\]
Using the monotonicity of norms~\eqref{eq:monotonicity_norms}, we recast
this inequality as one involving only a single power-mean,
\[
\left|\Omega\right|^{\frac{1}{4}}\langle\|c\|_4^4\rangle^{\frac{1}{4}}\left[\Big\langle\int_\Omega(\op\phi)^2d^\n
x\Big\rangle^{\frac{1}{2}}+\D\|\Delta\phi\|_\infty\langle\|c\|_4^4\rangle^{\frac{1}{2}}\right]\geq\left|\int_\Omega
s\phi d^\n x\right|.
\]
Therefore, we have the following inequality for
$m_4=\langle\|c\|_4^4\rangle^{\tfrac{1}{4}}$,
\[
m_4\left(q_0\left(\bm{v},D,\gamma\right)+\D\|\Delta\phi\|_\infty m_4^2\right)\geq\left|\Omega\right|^{-\frac{1}{4}}\left|\int_\Omega
s\phi d^\n x\right|,
\]
where
\[
q_0\left(\bm{v},D,\gamma\right)=\Big\langle\int_\Omega\left[\bm{v}\cdot\nabla\phi-\D\Delta\phi-\D\gamma\Delta^2\phi\right]^2d^\n
x\Big\rangle^{\frac{1}{2}}.
\]
Thus, we obtain a lower bound for the $m_4$ measure of composition
fluctuations,
\begin{equation}
m_4\geq m_4^{\mathrm{min}},\qquad \D\|\Delta\phi\|_\infty\left(m_4^{\mathrm{min}}\right)^3+q_0\left(\bm{v},s,D\right)m_4^{\mathrm{min}}-\left|\Omega\right|^{-\frac{1}{4}}\left|\int_{\Omega}
s\phi d^\n x\right|=0.
\label{eq:funky_variance}
\end{equation}
The cubic equation satisfied by $m_4^{\mathrm{min}}$ has a unique positive
root. 

To probe the asymptotic forms of~\eqref{eq:funky_variance}, we
rewrite the forcing terms $\bm{v}\left(\bm{x},t\right)$ and $s\left(\bm{x}\right)$
as an amplitude, multiplied by a dimensionless shape function.  Thus,
\begin{eqnarray*}
\bm{v}=V_0\tilde{\bm{v}}&,&\qquad V_0 = |\Omega|^{-\frac{1}{2}}\langle\|\bm{v}\|_2^2\rangle^{\frac{1}{2}},\\
s = S_0\tilde{s}&,&\qquad S_0 = |\Omega|^{-\frac{1}{2}}\|s\|_2,
\end{eqnarray*}
Then for a fixed value of $S_0$ and $D$, and $V_0\gg1$ (large stirring),
we have
$q_0\sim V_0\langle\int\left(\tilde{\bm{v}}\cdot\nabla\phi\right)^2d^{\n}x\rangle^{\frac{1}{2}}$,
and the lower bound $m_4^{\mathrm{min}}$ takes the form
\[
m_4^{\mathrm{min}}\sim\frac{S_0}{V_0}\frac{\left|\int_\Omega \tilde{s}\phi
d^{\n}x\right|}{\langle\int_\Omega\left(\tilde{\bm{v}}\cdot\nabla\phi\right)^2d^\n
x\rangle^{\frac{1}{2}}|\Omega|^{\frac{1}{4}}},\qquad V_0\gg1.
\]
On the other hand, for fixed $S_0$ and $V_0$, and $D\gg1$ (large diffusion),
the lower bound takes the form
\[
m_4^{\mathrm{min}}\sim
\frac{S_0}{D}\frac{\left|\langle\int_\Omega
\tilde{s}\phi d^\n x\rangle\right|}{\left[\int_\Omega\left(\Delta\phi+\gamma\Delta^2\phi\right)^2d^\n
x\right]^{\frac{1}{2}}|\Omega|^{\frac{1}{4}}},\qquad D\gg1.
\]

It is possible to obtain similar asymptotic expressions for $m_2$, by
constrained minimization of a functional of the composition.  Apart
from a volume factor, the asymptotic form of $m_2$ agrees exactly with
the asymptotic form of $m_4$ just obtained.  The functional to be
minimized is
\[
\Phi\left[c\right]=\tfrac{1}{2}\Big\langle\int_\Omega c^2d^{\n}x\Big\rangle-\lambda\Big\langle\int_\Omega\left(
c\op\phi+\D c^3\Delta\phi+s\phi\right)d^{\n}x\Big\rangle,
\]
where $\phi\left(\bm{x}\right)$ is a test function.  (This approach
was used in~\cite{Thiffeault2006} for the advection-diffusion
equation.)  Setting $\delta \Phi/\delta c=0$ gives
\begin{equation}
c=\frac{1-\sqrt{1-12\lambda^2\D\Delta\phi \op\phi}}{6\lambda \D\Delta\phi}.
\label{eq:c_sqrt}
\end{equation}
Evaluation of $\delta^2 \Phi/\delta c\,\delta c'$ shows that
Eq.~\eqref{eq:c_sqrt} produces a minimum of $\Phi\left[c\right]$.
Given the expression $\op=V_0\tilde{\bm{v}}\cdot\nabla-D\Delta-\gamma
D\Delta^2$, the minimum Eq.~\eqref{eq:c_sqrt} is $\lambda
V_0\tilde{\bm{v}}\cdot\nabla\phi$ at large $V_0$.  Substitution of
this expression into the constraint
$\langle\int_\Omega\left[c\op\phi+\D
  c^3\Delta\phi+s\phi\right]d^{\n}x\rangle=0$ gives
$\lambda=-\left(S_0/V_0^2\right)\left[\langle \int_\Omega
  \tilde{s}\phi d^{\n}x\rangle/
  \langle\int_\Omega\left(\tilde{\bm{v}}\cdot\nabla\phi\right)^2d^{\n}x\rangle\right]$,
and hence
\[
m_2^{\mathrm{min}}
\sim
\frac{S_0}{V_0}
\frac{\left|\langle\int_\Omega \tilde{s}\phi d^\n x\rangle\right|}
{\langle\int_\Omega\left(\tilde{\bm{v}}\cdot\nabla{c}\right)^2d^{\n}x\rangle^{\frac{1}{2}}},
\qquad V_0\gg1.
\]
For fixed $V_0$ and $S_0$ and large $D$, a similar calculation gives
\[
m_2^{\mathrm{min}}\sim
\frac{S_0}{D}\frac{\left|\langle\int_\Omega
\tilde{s}\phi d^\n x\rangle\right|}{\left[\int_\Omega\left(\Delta\phi+\gamma\Delta^2\phi\right)^2d^\n
x\right]^{\frac{1}{2}}},\qquad D\gg1.
\]

These expressions show that, apart from a volume factor, the lower
bounds on the $m_2$ and $m_4$ measures of composition fluctuations are
identical in the limits of high stirring strength or high diffusion.
In particular, the asymptotic expression
\[
m_2^{\mathrm{min}},\phantom{a}|\Omega|^{\frac{1}{4}}m_4^{\mathrm{min}}\sim\frac{S_0}{V_0}\frac{\left|\int_\Omega
\tilde{s}\phi d^{\n}x\right|}{\langle\int_\Omega\left(\tilde{\bm{v}}\cdot\nabla\phi\right)^2d^\n
x\rangle^{\frac{1}{2}}},\qquad \text{for large }V_0,
\]
indicates that if a flow can be found that saturates the lower bound
$m_{2,4}^{\mathrm{min}}$, the suppression of composition fluctuations
can be enhanced by a factor of $V_0^{-1}$ at large stirring
amplitudes.  Such a flow would then be an efficient way of mixing the
binary fluid.

\section{Scaling laws for $m_4$}
\label{sec:scaling}

In this section we investigate the dependence of the $m_4$ measure
of composition fluctuations on the parameters of the problem, namely the
stirring velocity $\bm{v}$, the source $s$, and the diffusion constant $D$.
 For simplicity, we shall restrict our interest to a certain class of flows,
 which enables us to compute long-time averages explicitly.

\begin{figure}
\subfigure[]{
  \scalebox{0.50}[0.50]{\includegraphics*[viewport=0 0 400 320]{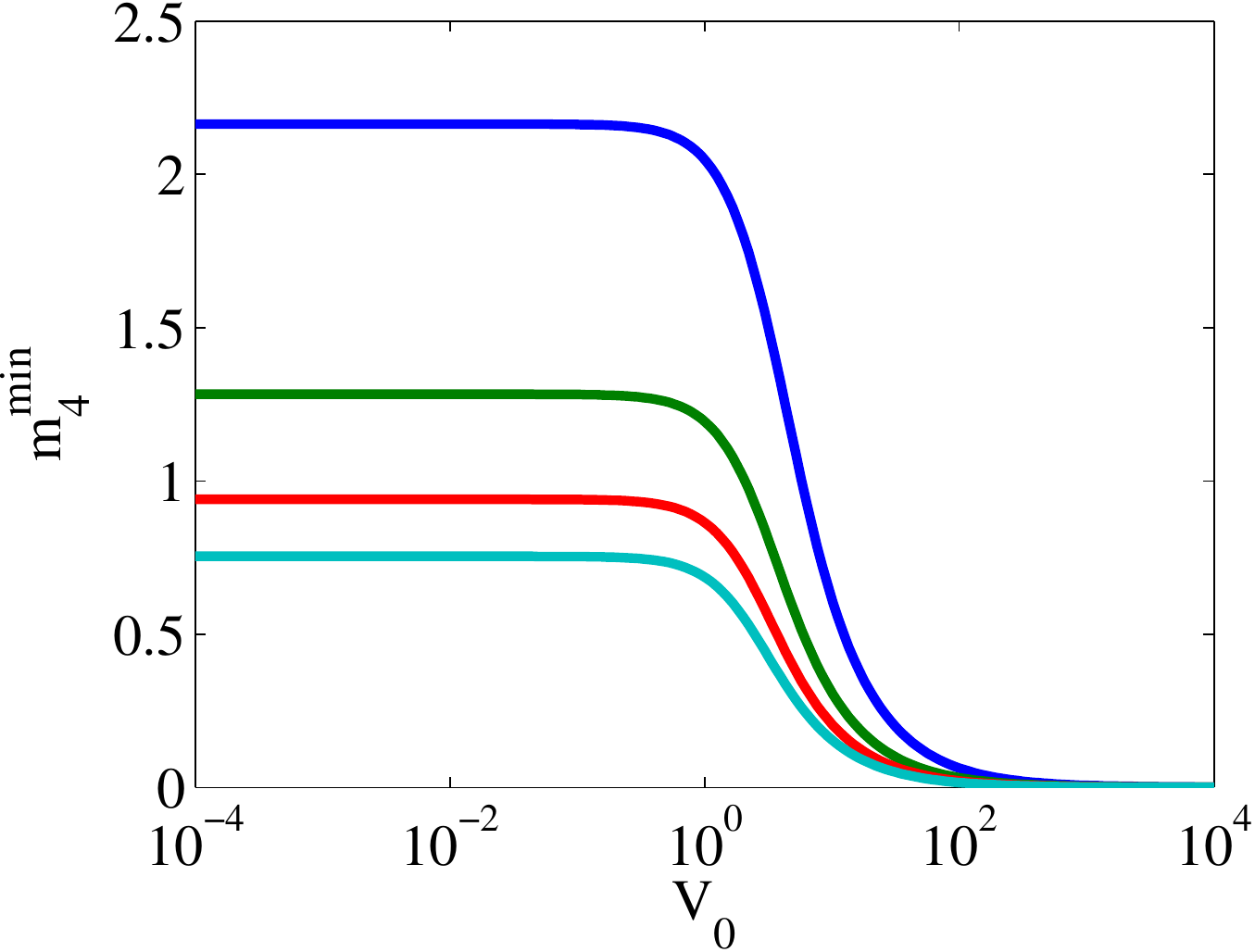}}
}
\subfigure[]{
  \scalebox{0.5}[0.50]{\includegraphics*[viewport=0 0 400 320]{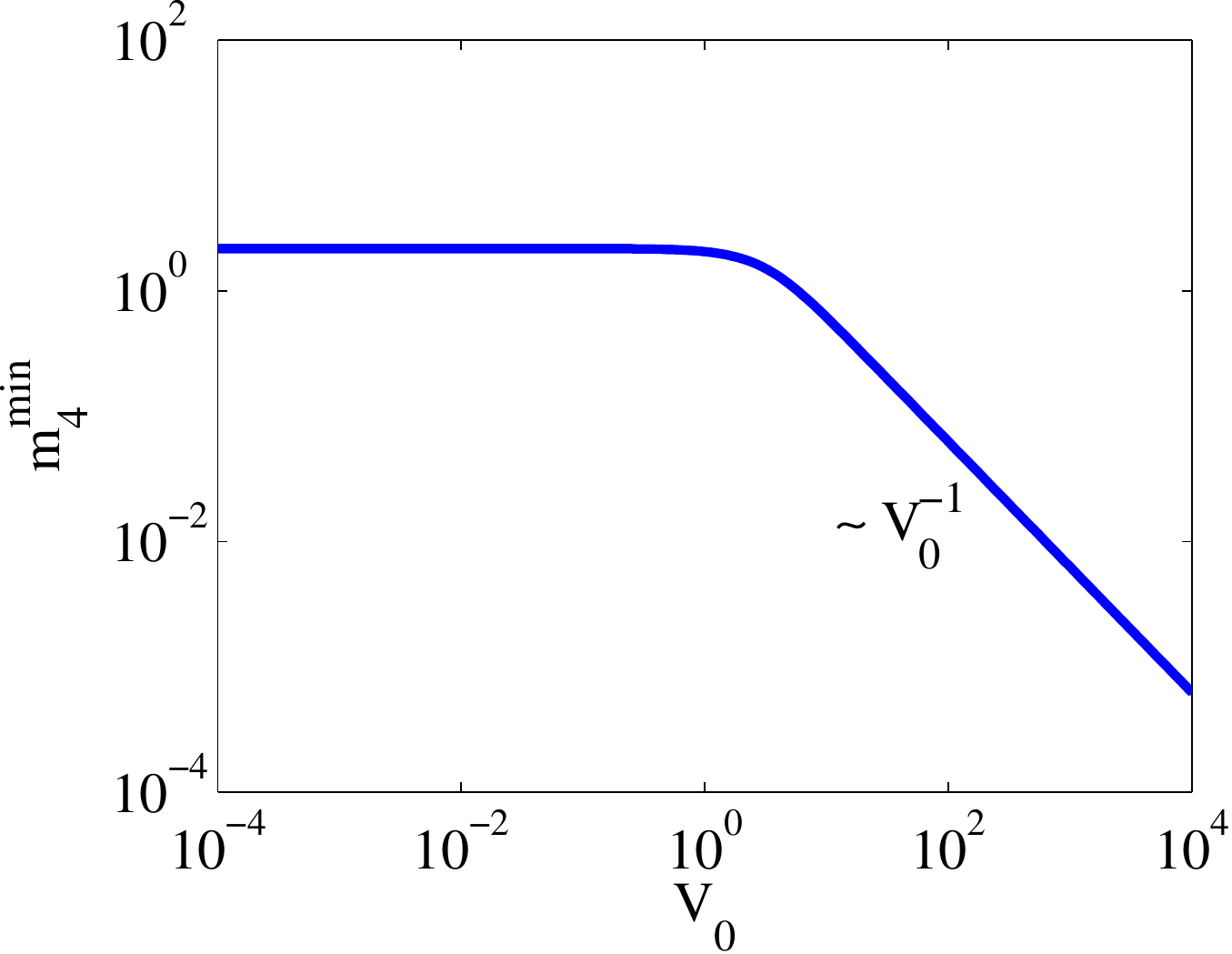}}
}
\caption{(a) The lower bound for $m_4^{\mathrm{min}}$ as a function of
  $V_0$ for a monochromatic source; (b) the dependence of $m_4^{\mathrm{min}}$
  on
  the velocity amplitude $V_0$.  In (a), the scale of the source
  variation decreases in integer multiples from
  $k_{\mathrm{s}}=2\pi/L$ in the uppermost curve, to
  $k_{\mathrm{s}}=8\pi/L$ in the lowermost curve, while in (b) the
  source scale is set to $2\pi/L$.  In both figures, we have set
  $D=S_0=1$.}
\label{fig:monochromatic}
\end{figure}

The lower bound for the $m_4$ measure of composition fluctuations is the
unique positive root of the polynomial
\begin{equation}
\D\|\Delta\phi\|_\infty\left(m_4^{\mathrm{min}}\right)^3+q_0\left(\bm{v},D,\gamma\right)m_4^{\mathrm{min}}-\left|\Omega\right|^{-\frac{1}{4}}\left|\int_{\Omega}
s\phi\, d^\n x\right|=0,
\label{eq:lower_bd_monochromatic}
\end{equation}
where $\phi\left(\bm{x}\right)$ is a test function and
\[
q_0\left(\bm{v},D,\gamma\right)=\Big\langle\int_\Omega\left[\bm{v}\cdot\nabla\phi-\D\Delta\phi-\D\gamma\Delta^2\phi\right]^2d^\n
x\Big\rangle^{\frac{1}{2}}.
\]
Following~\cite{DoeringThiffeault2006,Shaw2007}, we specialize to
velocity fields whose time average has the following properties,
\begin{equation}
\langle v_i\left(\bm{x},\cdot\right)\rangle=0,
\qquad\langle v_i\left(\bm{x},\cdot\right)v_j\left(\bm{x},\cdot\right)\rangle=\frac{V_0^2}{\n}\delta_{ij}.
\label{eq:statistical_hit}
\end{equation}
The flow $\bm{v}\left(\bm{x},t\right)$ is defined on the $n$-torus
$\left[0,L\right]^n$.  A statistically homogeneous and isotropic
turbulent velocity field automatically satisfies the
relations~\eqref{eq:statistical_hit}, although it is not necessary for
$\bm{v}$ to be of this type.  The source we consider is monochromatic
(that is, it contains contains a single spatial scale) and varies in a
single direction,
\begin{equation}
s=\sqrt{2}S_0\sin\left(k_{\mathrm{s}}x\right).
\label{eq:monochromatic}
\end{equation}
Our choice of velocity field makes the evaluation of
$q_0\left(\bm{v},D,\gamma\right)$ particularly easy:
\[
q_0\left(\bm{v},D,\gamma\right)=\left[\frac{V_0^2}{\n}\int_{\left[0,L\right]^n}\left|\nabla\phi\right|^2d^{\n}x+D^2\int_{\left[0,L\right]^n}\left(\Delta\phi+\gamma\Delta^2\phi\right)^2d^{\n}x\right]^{\frac{1}{2}}.
\]
In studies of the advection-diffusion equation~\cite{Thiffeault2006}, it
is possible to find an explicit test function $\phi$ that sharpens the lower
bound on $m_2$.  The procedure for doing this depends on the linearity of
the equation.
Here, this is not possible, and for simplicity we set $\phi=s$.  This
choice of $\phi$ certainly gives a lower bound for the $m_4$ measure
of composition fluctuations, with the added advantage of enabling
explicit computations.  Having specified the coefficients of the
polynomial in Eq.~\eqref{eq:lower_bd_monochromatic} completely, we
extract the positive root of this equation, and find the lower bound
$m_4^{\mathrm{max}}$, as a function of $V_0$.  The results of this
procedure are shown in Fig.~\ref{fig:monochromatic}.

\begin{figure}[htb]
 \scalebox{0.50}[0.50]{\includegraphics*[viewport=0 0 400 320]{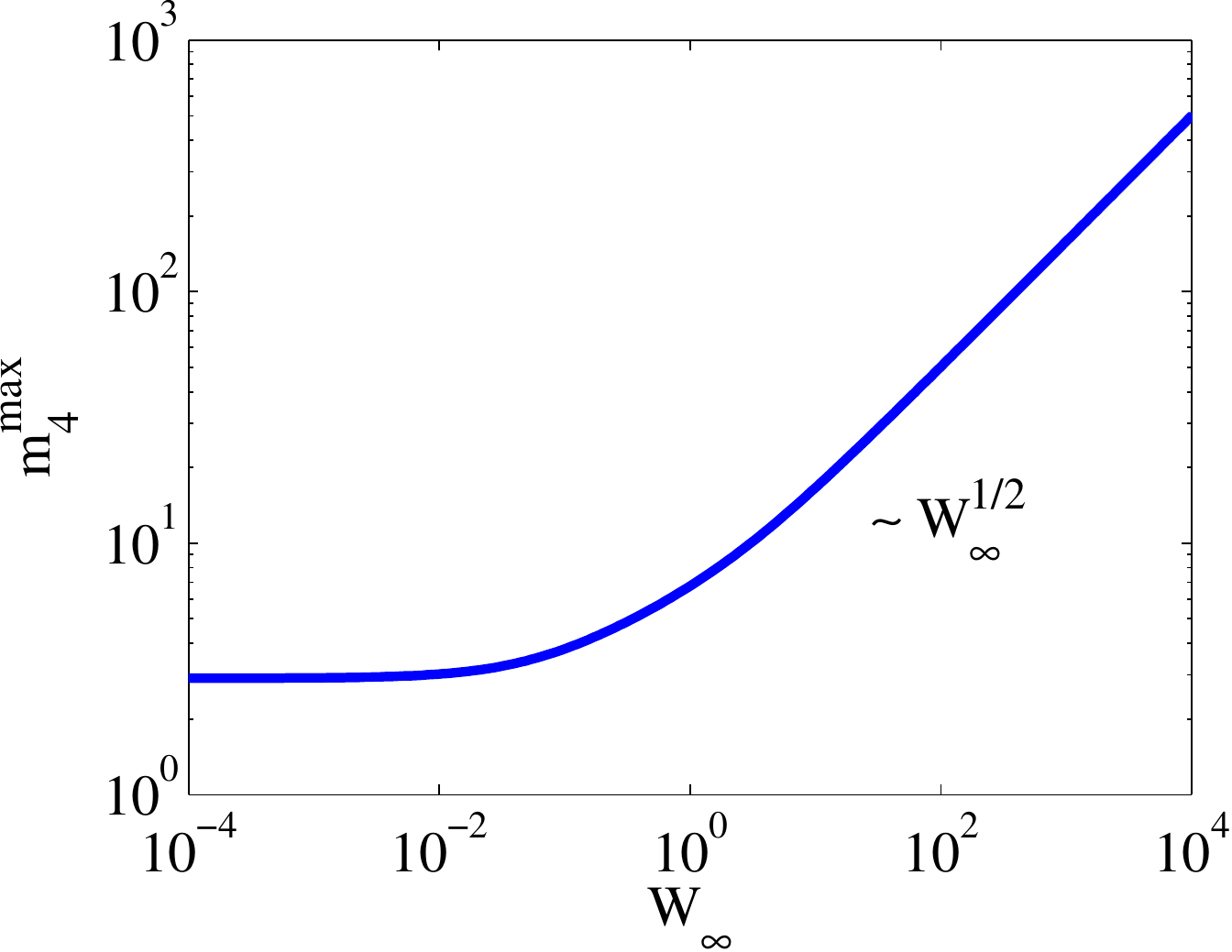}}
 \caption{The dependence of the upper bound $m_4^{\mathrm{max}}$ on
   the maximum rate-of-strain $W_{\infty}$.  The source affects the upper
   bound only through its square mean
   $S_0=|\Omega|^{-\frac{1}{2}}\|s\|_2$.}
\label{fig:upper_bd}
\end{figure}

Let us examine briefly the scaling of the upper bound $m_4^{\mathrm{max}}$
with the problem parameters.  The upper bound satisfies the polynomial equation
\begin{multline}
\left(m_{4}^{\mathrm{max}}\right)^4=\left|\Omega\right|
\\
+\frac{2\left|\Omega\right|^{\frac{1}{4}}}{D}\left(\frac{L}{2\pi}\right)^2m_{4}^{\mathrm{max}}\left[\left(S_0|\Omega|^{\frac{1}{2}}+\n\left|\Omega\right|^{\frac{1}{4}}\Deform_\infty{m_{4}}^{\mathrm{max}}\right)^2+{4\n
D\left|\Omega\right|^{\frac{1}{2}}}\left(\frac{2\pi}{L}\right)^2\Deform_\infty\left({m_{4}}^{\mathrm{max}}\right)^2\right]^{\frac{1}{2}},
\label{eq:upper_bd_monochromatic}
\end{multline} 
which depends only on the diffusion $D$, the source amplitude $S_0$,
and the maximum rate-of-strain $\Deform_\infty$.  For $\Deform_\infty$
large, the flow-dependence of the upper bound is
$m_4^{\mathrm{max}}\sim\Deform_\infty^{\frac{1}{2}}$.  This dependence
is verified by obtaining the positive root of
Eq.~\eqref{eq:upper_bd_monochromatic}, which is a function of
$W_\infty$.  The results are shown in Fig.~\ref{fig:upper_bd}.

In this section we have investigated the parametric dependence of the theoretical
bounds $m_4^{\text{max, min}}$, for flows with the properties in Eq.~\eqref{eq:statistical_hit}.
We note that for any nonzero source, the lower bound $m_4^{\mathrm{min}}$
is nonzero, meaning that no matter how hard one stirs, there will always
be some inhomogeneity in the fluid, and this is in fact true for any flow.
 However, the number $m_4^{\mathrm{min}}$ tells us how much homogeneity we
 can achieve and is therefore a yardstick for stirring protocols.  We use
 this yardstick to test model flows in the next section.

\section{Numerical Simulations}
\label{sec:numerics}

In this section we solve Eq.~\eqref{eq:ch} numerically for two
flows, and verify the bounds obtained in Secs.~\ref{sec:existence}--\ref{sec:scaling}.
 We use the sinusoidal source term in Eq.~\eqref{eq:monochromatic} with periodic
 boundary conditions, and the source scale $k_{\mathrm{s}}$ therefore takes
 the form $\left(2\pi/L\right)j$, where $L$ is the box size and $j$ is an
 integer.  We specialize to two dimensions and study two standard flows that
 are used in the analysis of mixing: the random-phase sine flow~\cite{lattice_PH2,Antonsen1996,Neufeld_filaments,lattice_PH1,
 Thiffeault2004}, and the constant flow~\cite{Thiffeault2007}.

\subsection*{Random-phase sine flow}

The random-phase sine flow is the time-dependent two-dimensional
flow
\begin{equation}
\begin{split}
  v_x\left(x,y,t\right) &= \sqrt{2}\,V_0\sin\left(k_{\mathrm{v}}y+\phi_j\right),\qquad
  v_y=0,\qquad j\tau\leq t<\left(j+\tfrac{1}{2}\right)\tau,\\
  v_y\left(x,y,t\right) &= \sqrt{2}\,V_0\sin\left(k_{\mathrm{v}}x+\psi_j\right),\qquad
  v_x=0,\qquad \left(j+\tfrac{1}{2}\right)\tau \leq t<\left(j+1\right)\tau,
\end{split}
\label{eq:sineflow}
\end{equation}
where $\phi_j$ and $\psi_j$ are phases that are randomized once during each
flow period $\tau$, and where the integer $j$ labels the period.  The flow
is defined
on the two-dimensional torus $\left[0,L\right]^2$.  The time average of this
velocity field has the properties listed in Eq.~\eqref{eq:statistical_hit}.
Because of its simplicity, the sine flow is a popular testbed for studying
chaotic mixing~\cite{lattice_PH2,Antonsen1996,Neufeld_filaments,lattice_PH1,%
Thiffeault2004}.

We solve Eq.~\eqref{eq:ch} with the flow in Eq.~\eqref{eq:sineflow}
using an operator splitting: the advection step is carried out using
the lattice method of Pierrehumbert~\cite{lattice_PH1, lattice_PH2},
and the subsequent Cahn--Hilliard and source steps are implemented
simultaneously using a spectral method~\cite{Zhu_numerics}.  The
nondimensionalization outlined in Sec.~\ref{sec:model} is appropriate
here: the unit of time $\Time$ is identified with the is the flow
period $\tau$, and the unit of length is the box size $L$.  The
control parameters in the problem are the dimensionless velocity
$V_0$, the dimensionless diffusion $D$, and the dimensionless source
amplitude $S_0=|\Omega|^{-\frac{1}{2}}\|s\|_2$.  We use $V_0$ as a
measure of stirring intensity and fix the other parameters in what
follows.  The flow we choose is chaotic at all stirring amplitudes and
given our choice of scaling, has Lyapunov exponent~\cite{ONaraigh2007}
\[
\lambda\sim0.236\,{V_0}^2,\phantom{a}V_0\ll1;\qquad\lambda\sim\log\left(\frac{V_0^2}{2}\right),\phantom{a}V_0\gg1.
\]
The lattice method with its splitting of the advection and diffusion
steps, is effective only when diffusion is slow compared to advection,
that is, $\Time/\Time_D\ll1$.  We therefore set $D=10^{-5}$, with
$\tau=L=1$.  A numerical experiment with $V_0=0$ shows that
$S_0=5\times10^{-4}$ gives rise to a morphology that is qualitatively
different from the sourceless case, and we therefore work with
this source amplitude.  Finally, following standard
practice~\cite{Berti2005, chaos_Berthier}, we choose $\gamma\sim\Delta
x^2$, the gridsize.

 Using these new
 scaling rules, and the identity $\Deform_\infty=\sqrt{2}V_0k_{\mathrm{v}}$
 for the sine
 flow, we recall the large-stirring forms of $m_4^{\mathrm{max,min}}$,
For $V_0\gg\D$, the lower bound has the form
\begin{equation}
m_4^{\mathrm{min}}\sim\frac{S_0}{V_0}\frac{\int_\Omega\sin\left(k_{\mathrm{s}}x\right)\phi
d^2x}{\left[\int_\Omega\left|\nabla\phi\right|^2d^2x\right]^{\frac{1}{2}}},
\label{eq:lower_bd_asymp}
\end{equation}
with the power-law relationship $m_4^{\mathrm{min}}\sim V_0^{-1}$,
while for small $V_0\gg1$ the upper bound has the form
\begin{equation}
m_4^{\mathrm{max}}\sim\left(\frac{2^{\frac{1}{2}}k_{\mathrm{v}}}{\pi^2D}\right)^{\frac{1}{2}}V_0^{\frac{1}{2}}.
\label{eq:upper_bd_asymp}
\end{equation}
These scaling results are identical to those for the advection-diffusion
problem~\cite{Thiffeault2004}.

Before studying the case with flow, we integrate Eq.~\eqref{eq:ch}
without flow, to verify the effect of the source.
\begin{figure}[htb]
\subfigure[]{
  \scalebox{0.23}[0.23]{\includegraphics*[viewport=0 0 450 380]{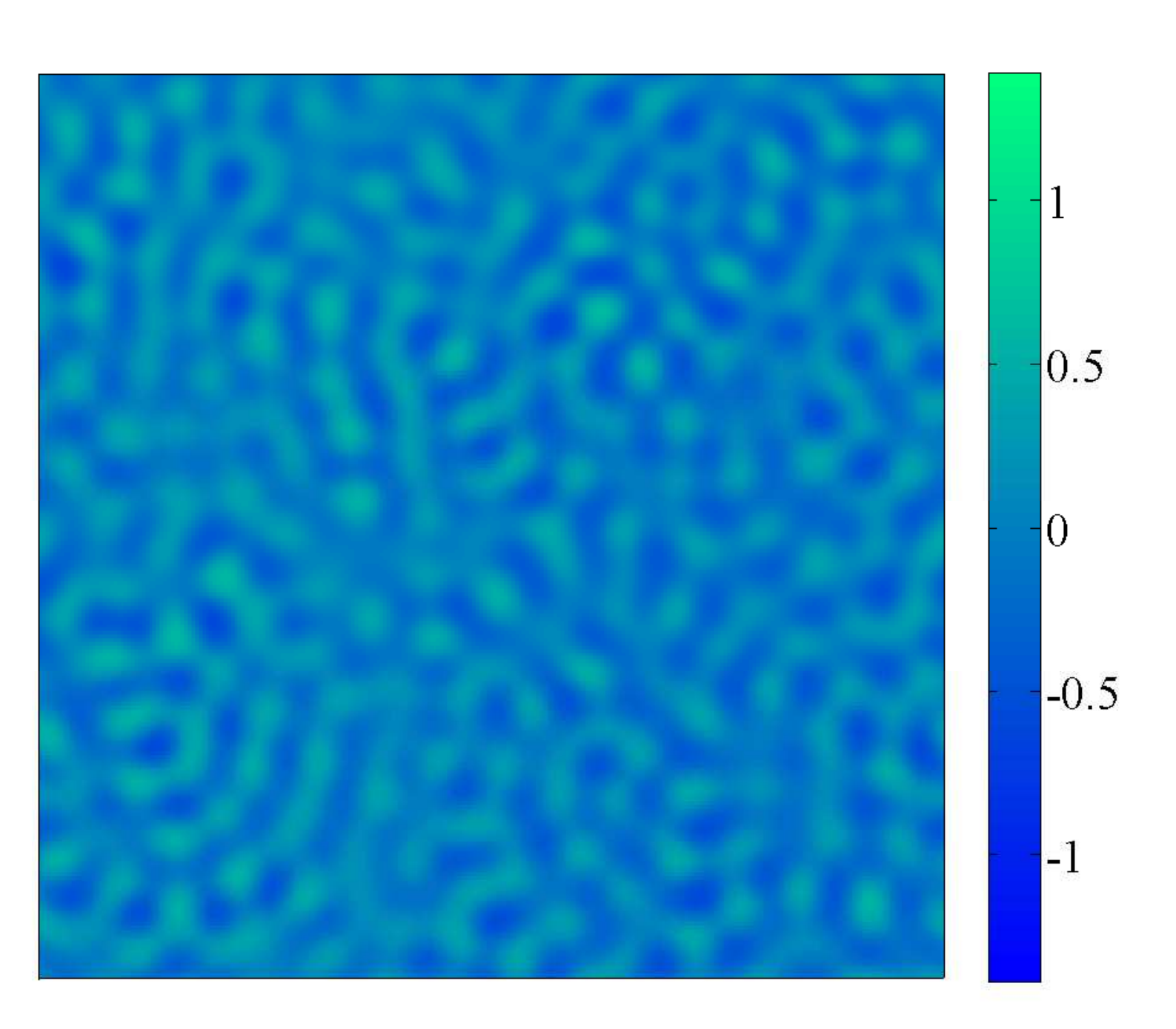}}
}
\subfigure[]{
  \scalebox{0.23}[0.23]{\includegraphics*[viewport=0 0 380 380]{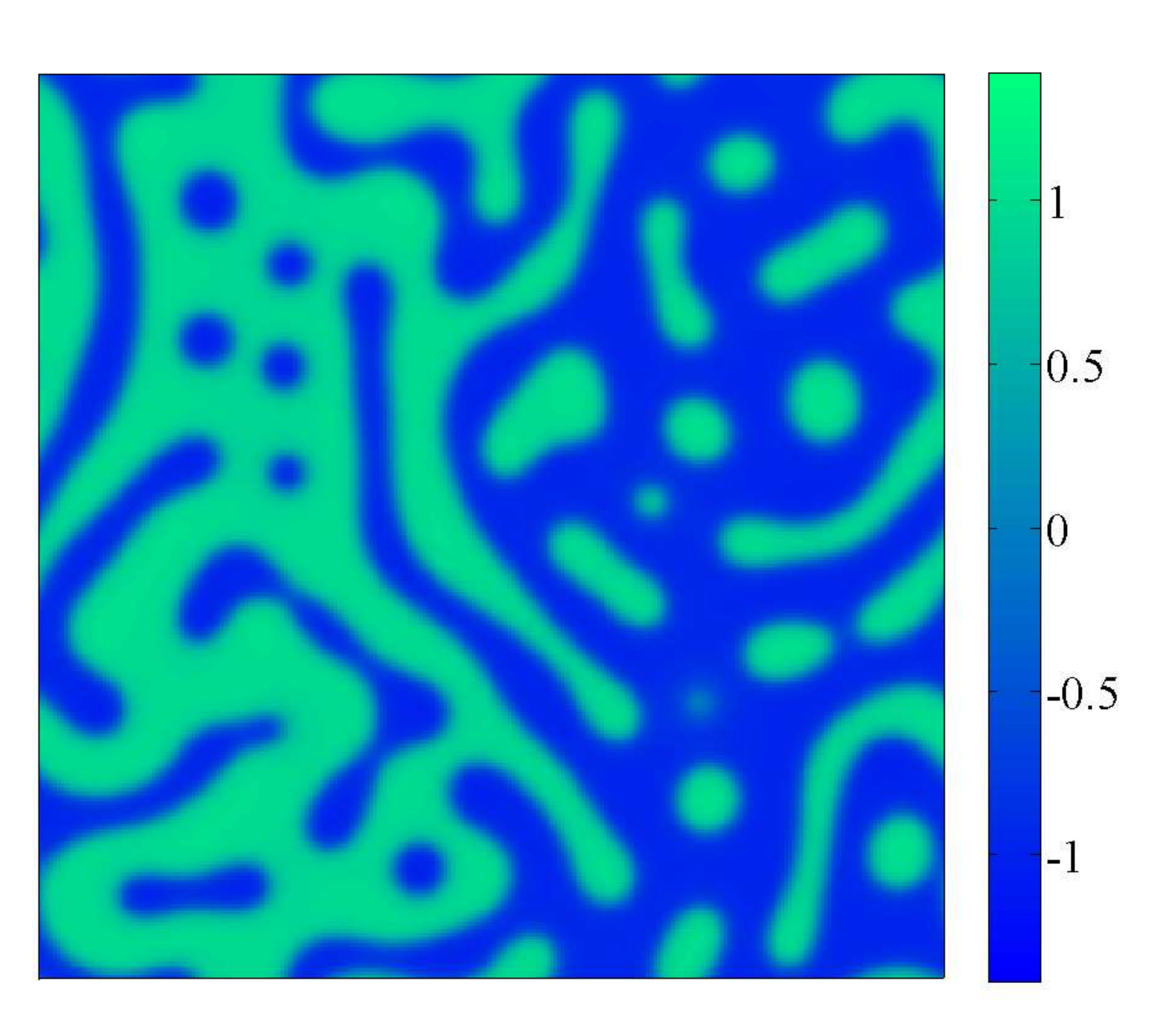}}
}
\subfigure[]{
  \scalebox{0.23}[0.23]{\includegraphics*[viewport=0 0 380 380]{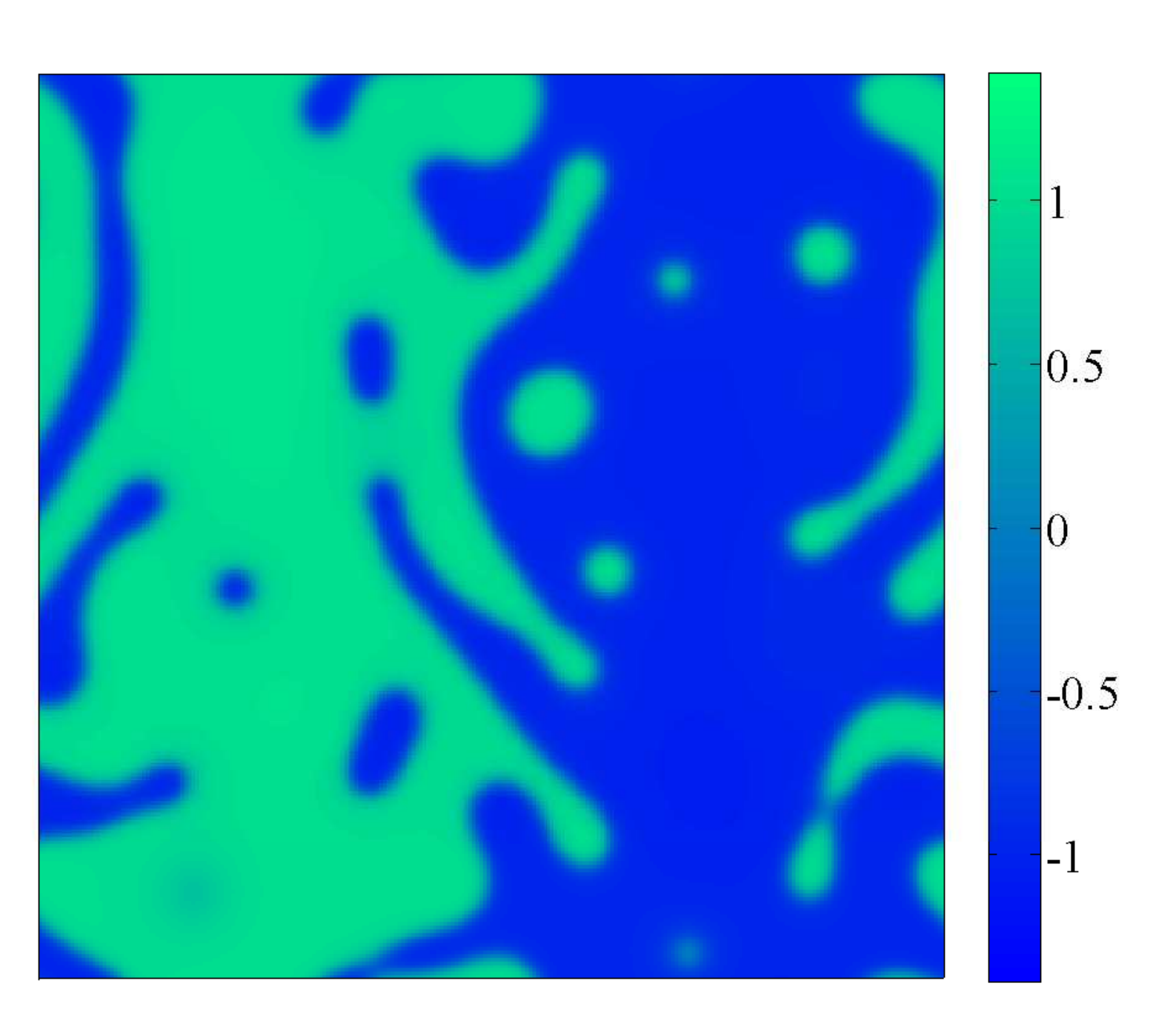}}
}
\subfigure[]{
  \scalebox{0.23}[0.23]{\includegraphics*[viewport=0 0 380 380]{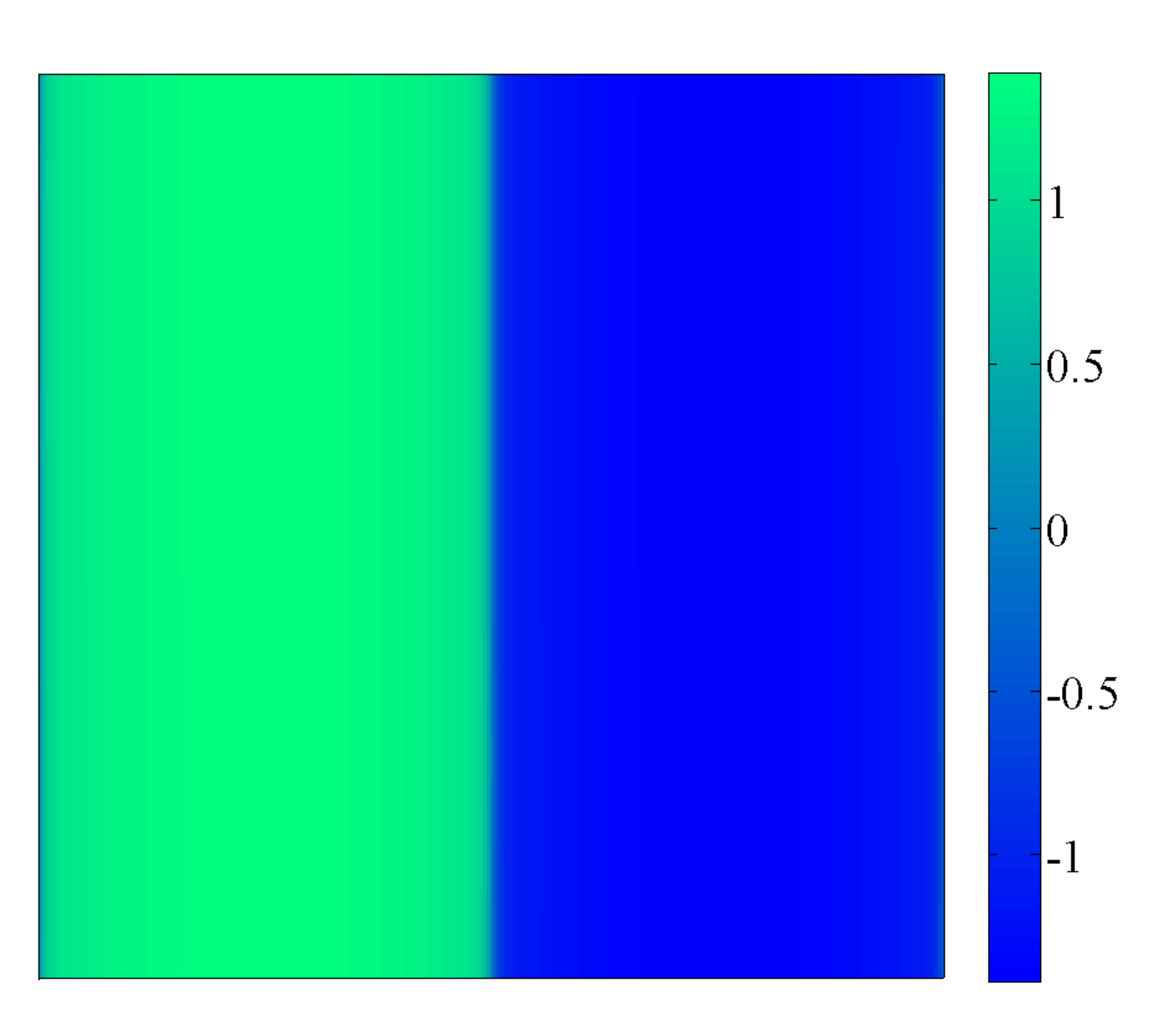}}
}
\subfigure[]{
  \scalebox{0.5}[0.5]{\includegraphics*[viewport=0 0 400 310]{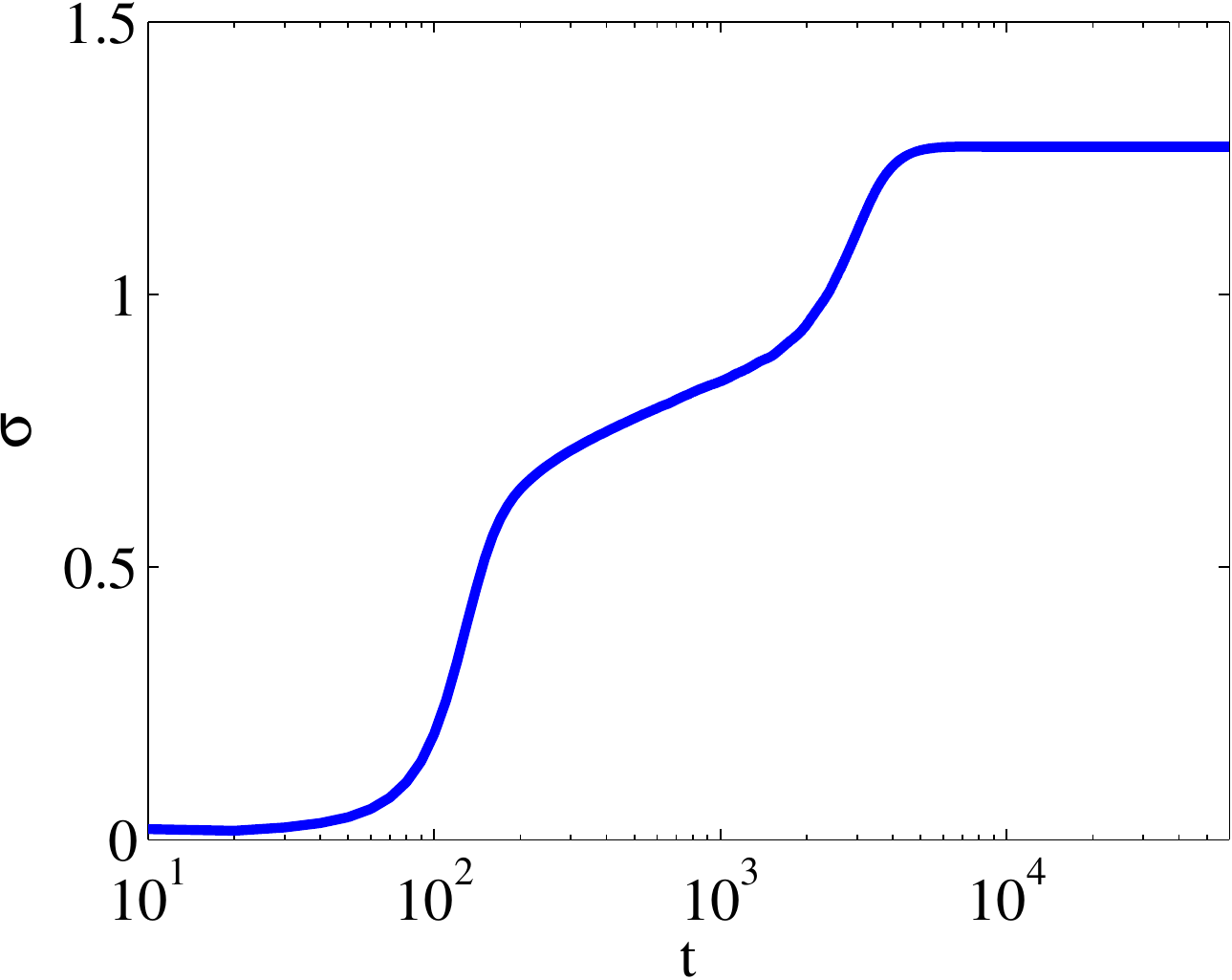}}
}
\caption{The composition of the binary fluid for $S_0 = 5\times10^{-4}$ and
(a) $t=100$; (b) $t=1000$; (c) $t=2000$; (d) $t=8000$.  A steady state is
reached in (d), evidenced by the time dependence of $m_4$ in
(e), where $m_4$ is constant for $t\gg1$.}
\label{fig:no_flow}
\end{figure}
For a sufficiently large source amplitude, the composition phase-separates
and forms domains rich in either binary fluid component.  These domains are
aligned with variations in the source.  A steady-state
is reached and $m_4$ attains a constant value, as seen in Fig.~\ref{fig:no_flow}.
 On the other hand, for small source amplitudes, we have verified that that
 the domains do not align with the source, and their growth
 does not saturate.  We do not consider this case here, since we are interested
 in sources that qualitatively alter the phase separation.  These different
 regimes are discussed in~\cite{Krekhov2004}.

\begin{figure}[htb]
\subfigure[]{
  \scalebox{0.25}[0.25]{\includegraphics*[viewport=0 0 450 400]{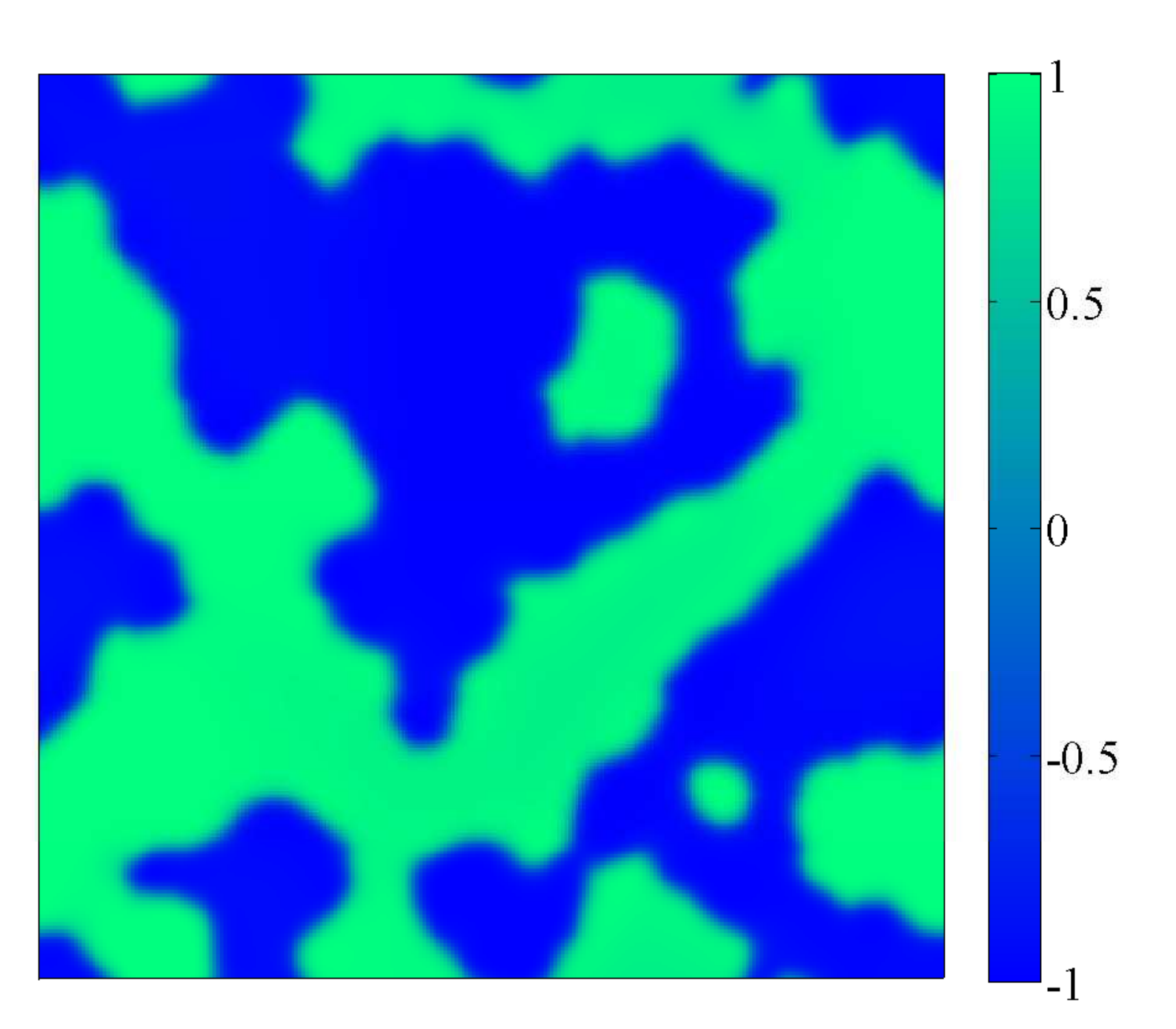}}
}
\subfigure[]{
  \scalebox{0.25}[0.25]{\includegraphics*[viewport=0 0 450 400]{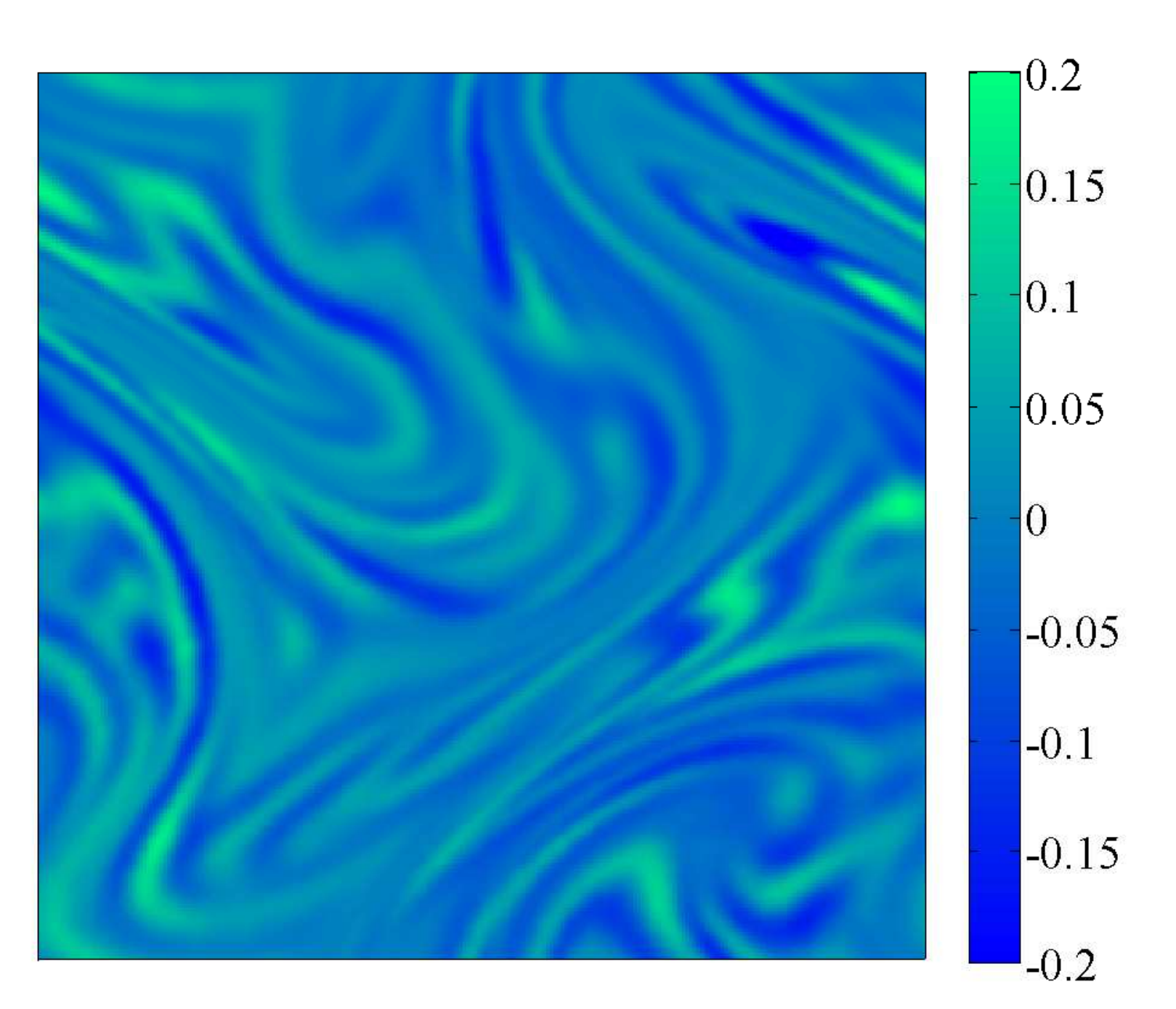}}
}
\subfigure[]{
  \scalebox{0.25}[0.25]{\includegraphics*[viewport=0 0 450 400]{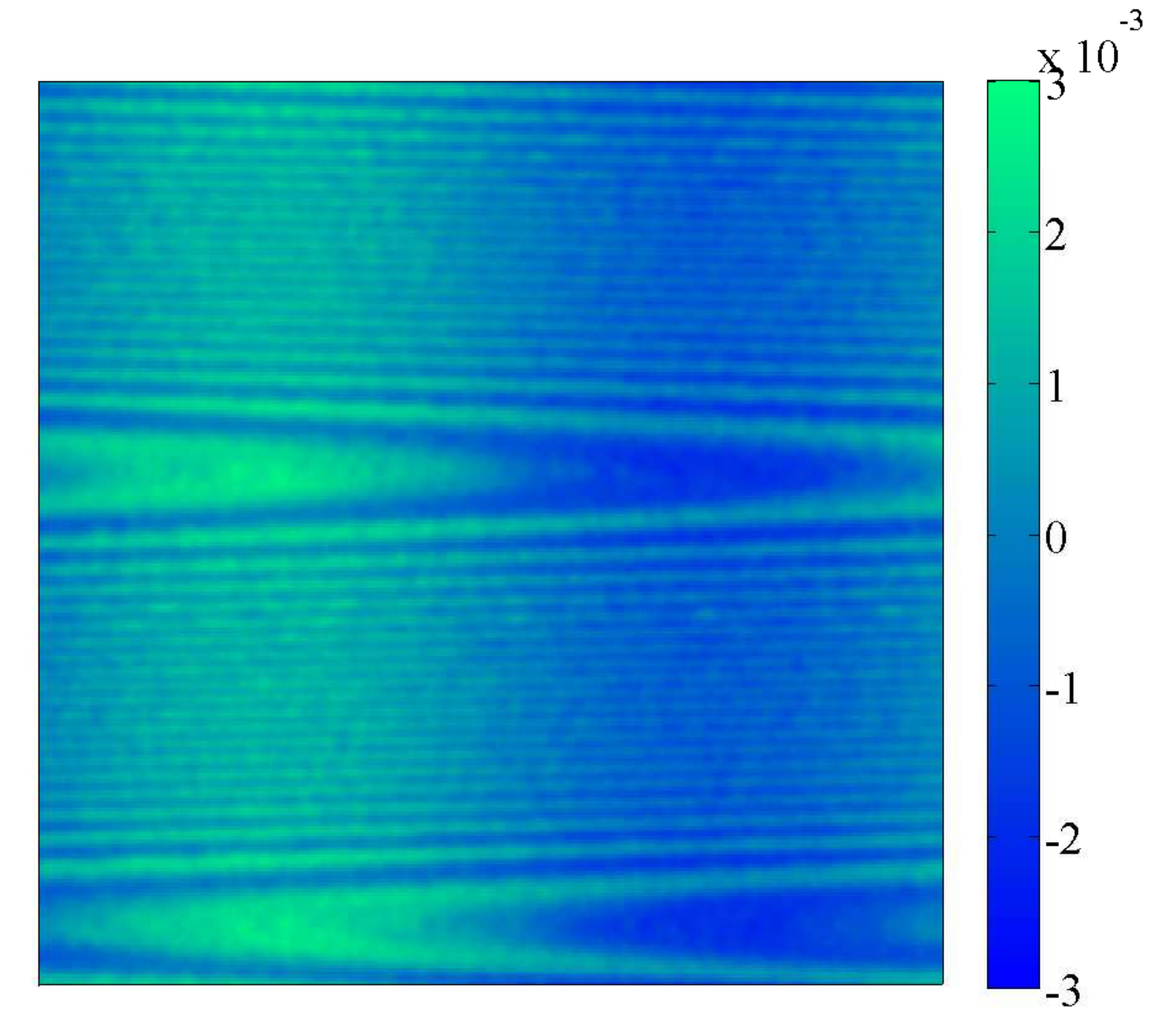}}
}
\caption{A snapshot of the steady-state composition for (a)
$V_0=0.001$; (b) $V_0=0.1$; (c) $V_0=10$.  In (a) domain growth
is arrested, in (b) the domains are destroyed and the binary fluid mixes,
while in (c) $m_4$ measure of composition fluctuations is minimized, and
the source structure is visible.}
\label{fig:C_flow}
\end{figure}

We consider the case with flow by varying $V_0$, and find results that
are similar to those found in~\cite{ONaraigh2007}, for the same stirring
mechanism without sources.  For all values of $V_0$, the composition
reaches a steady state, in which $\|c\|_4$, the pre-averaged form of $m_4$,
fluctuates around
a constant value.
For small values of $V_0$, the domain growth
is arrested due to a balance between the advection and phase-separation
terms in the equation, while for moderate values of $V_0$, the domains
are broken up and a mixed state is obtained.  At large values of $V_0$,
the $m_4$ measure of composition fluctuations saturates: further
increases in $V_0$ do not produce further decreases in $m_4$.
 At these large values of $V_0$, the source structure is visible in snapshots
 of the composition, as evidenced by Fig.~\ref{fig:C_flow}.

\begin{figure}
\subfigure[]{
  \scalebox{0.50}[0.50]{\includegraphics*[viewport=0 0 400 320]{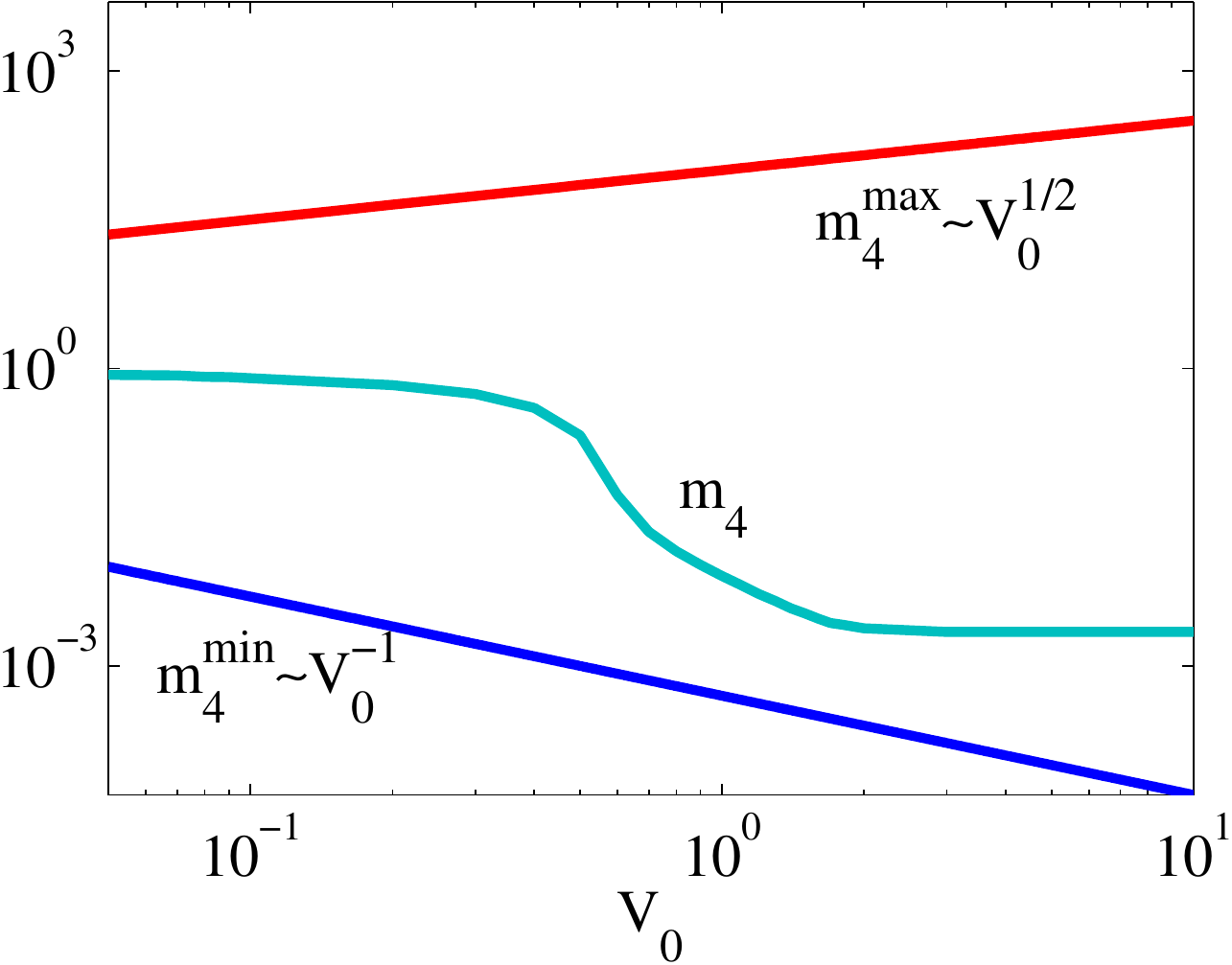}}
}
\subfigure[]{
  \scalebox{0.50}[0.50]{\includegraphics*[viewport=0 0 400 320]{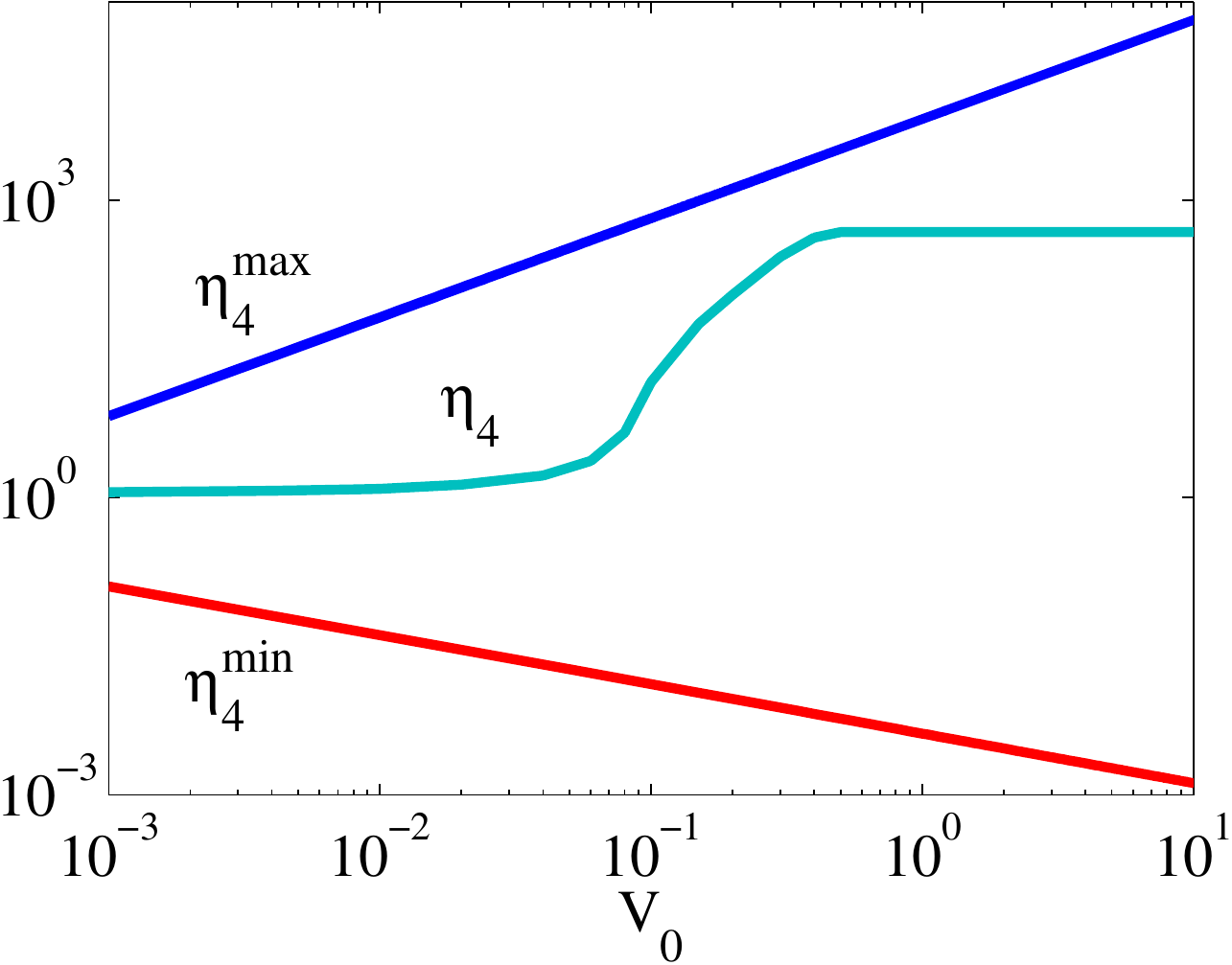}}
}
\caption{(a) The $m_4$ measure of composition fluctuations or mixing for
the
sine flow, as a function of the stirring parameter $V_0$.  The values of
$D$ and $S_0$ are given in the text.  The upper and lower bounds are shown
for comparison; (b) The mixing enhancement $\eta_4$ for the sine flow, with
the upper and lower bounds shown for comparison.}
\label{fig:min_max_sineflow}
\end{figure}

We investigate the dependence of composition fluctuations
on the stirring strength $V_0$, and show the results in Fig.~\ref{fig:min_max_sineflow}.
 The theoretical upper and lower bounds on $m_4$ depend on $V_0$ and are
 obtained as as roots of Eqs.~\eqref{eq:lower_bd_monochromatic} and~\eqref{eq:upper_bd_monochromatic}.
  In the limit of large $V_0$, these bounds have
  power-law behaviour, with $m_4^{\mathrm{max}}\sim V_0^{\frac{1}{2}}$
  and $m_4^{\mathrm{min}}\sim V_0^{-1}$, as demonstrated by
  Eqs.~\eqref{eq:lower_bd_asymp} and~\eqref{eq:upper_bd_asymp}.  The numerical
  values of $m_4$ are indeed bounded by these limiting values,
  although the $V_0$-dependence is not a power law.  Instead, the function
  $m_4\left(V_0\right)$
is a non-increasing function, with a sharp drop occurring in a small
range of $V_0$-values.  Thus, the fluid becomes more homogeneous with
increasing $V_0$.  We discuss the effect of stirring on the
inhomogeneity of the fluid by introducing the notion of mixing
enhancement.

We measure the ability of a given stirring protocol to suppress
composition fluctuations by the mixing enhancement.  Similar ideas are
often applied to the advection-diffusion
equation~\cite{DoeringThiffeault2006,Shaw2007,
  Thiffeault2004,Thiffeault2006,Thiffeault2007}.  We define the
dimensionless mixing enhancement
\[
\eta_p\equiv\frac{m_p^{\mathrm{min}}\left(V_0=0\right)}{m_p\left(V_0\right)}.
\]
For a given flow, the number $\eta_p$ quantifies the flow's ability to
suppress composition fluctuations.  In a well-mixed flow, the local
deviation of $c\left(\bm{x},t\right)$ away from the mean will be
small; a small $m_p$-value is a signature of a well-mixed flow.  We
are therefore justified in calling $\eta_p$ the mixing enhancement.
We obtain some control over the mixing enhancement $\eta_4$ from the
inequalities of Secs.~\ref{sec:existence} and~\ref{sec:lower_bds}.
Based on these inequalities, the mixing enhancement is bounded above
and below,
\[
\eta_4^{\mathrm{min}}\equiv\frac{m_4^{\mathrm{min}}\left(V_0=0\right)}{m_4^{\mathrm{max}}\left(V_0\right)}
\leq\eta_4
\leq\frac{m_4^{\mathrm{min}}\left(V_0=0\right)}{m_4^{\mathrm{min}}\left(V_0\right)}\equiv\eta_4^{\mathrm{max}}.
\]
We have plotted the upper and lower bounds on the mixing enhancement for the
case of monochromatic sources in Fig.~\ref{fig:min_max_sineflow}(b).  The
maximum enhancement
always exceeds unity in this case, which implies the possibility of
finding stirring protocols that homogenize the fluid.  On the other
hand, the minimum enhancement is less than unity, which indicates the
possibility of finding stirring protocols that actually amplify
composition fluctuations, and this amplification depends weakly on the
maximum rate-of-strain $\Deform_\infty$.  This latter case is not
surprising, given that a uniform shear flow causes the domains of the
Cahn--Hilliard fluid to align, rather than to break up.  The sine-flow
enhancement is a non-decreasing function of the stirring parameter $V_0$.
At small values of $V_0$, small increases in the vigor of stirring
lead to small small increases in the mixing enhancement.  There is a
window of intermediate $V_0$-values for which the mixing enhancement
increases sharply with increasing $V_0$.  At higher values of $V_0$,
the efficiency saturates, so that further increases in the vigor of
stirring have no effect on composition fluctuations.  The saturation
is due to finite-size effects: the sine flow wraps filaments of fluid
around the torus as in Fig.~\ref{fig:C_flow}(c).

\subsection*{Constant flow}

We study the flow $\left(v_x,v_y\right)=\left(0,V_0\right)$,
where $V_0$ is a constant.  We choose a nondimensionalization that is set
by the diffusion time $T_D=L^2/D$, and obtain the following parametric version
of Eq.~\eqref{eq:ch},
\begin{equation}
\frac{\partial c}{\partial t'}+V_0'\frac{\partial c}{\partial y'}=\Delta'\left(c^3-c-\gamma'\Delta'
c\right)+\sqrt{2}S_0'\sin\left(k_{\mathrm{s}}'x'\right),
\label{eq:const_flow}
\end{equation}
where $V_0'=LV_0/D$, $\gamma'=\gamma/L^2$, and $S_0=S_0'L^2/D$.  We
immediately drop the primes from Eq.~\eqref{eq:const_flow}.  We fix
$\gamma$ and $S_0$ and vary the flow strength $V_0$.  This flow does
not satisfy the time-correlation relations~\eqref{eq:statistical_hit},
although the maximum rate-of-strain has the simple form
$\Deform_\infty=0$.  Using this information, and the test function
$\phi=s\left(x\right)$, the
\begin{figure}[htb]
\subfigure[]{
  \scalebox{0.25}[0.25]{\includegraphics*[viewport=0 0 450 400]{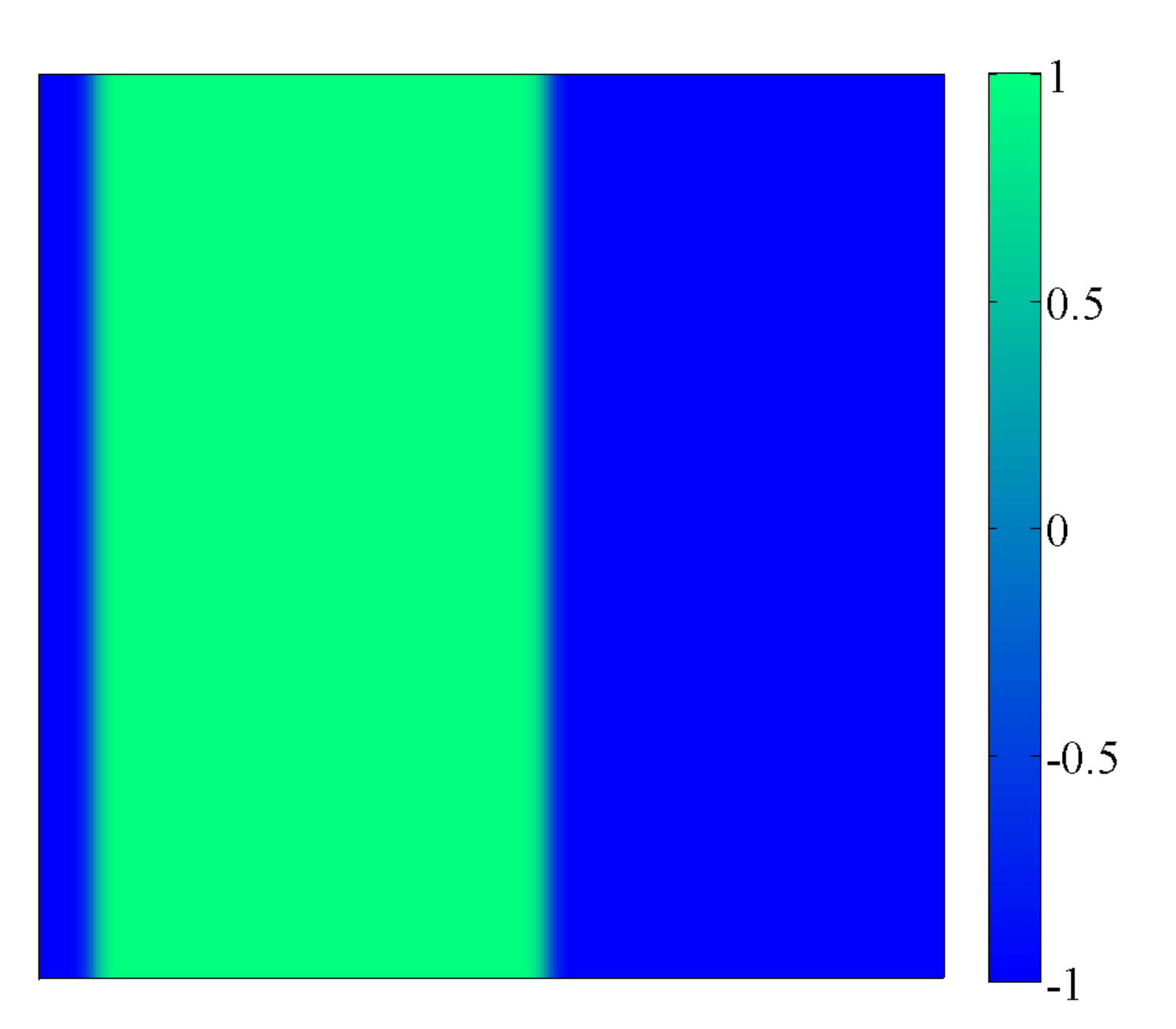}}
}
\subfigure[]{
  \scalebox{0.25}[0.25]{\includegraphics*[viewport=0 0 380 400]{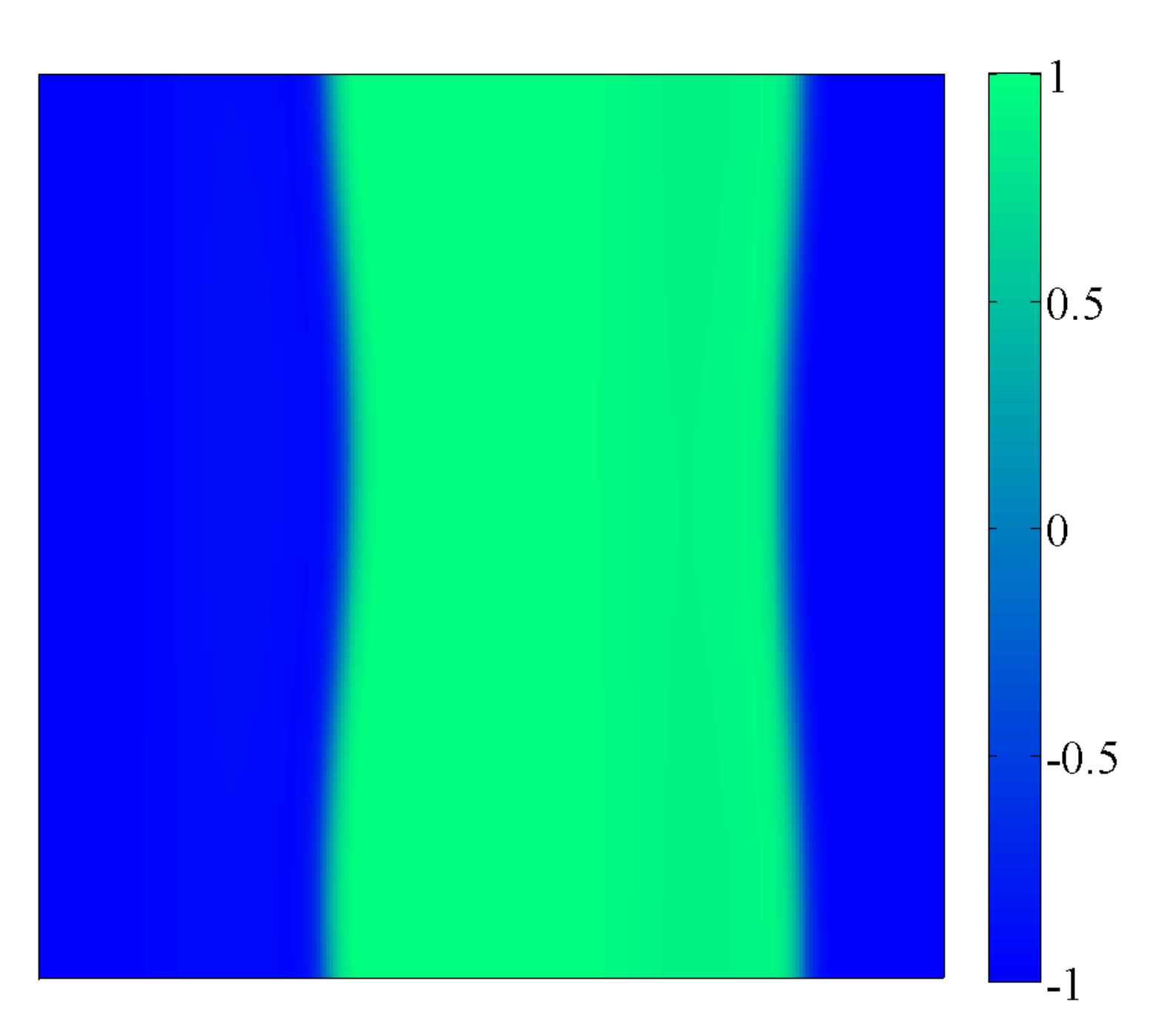}}
}
\subfigure[]{
  \scalebox{0.25}[0.25]{\includegraphics*[viewport=0 0 380 400]{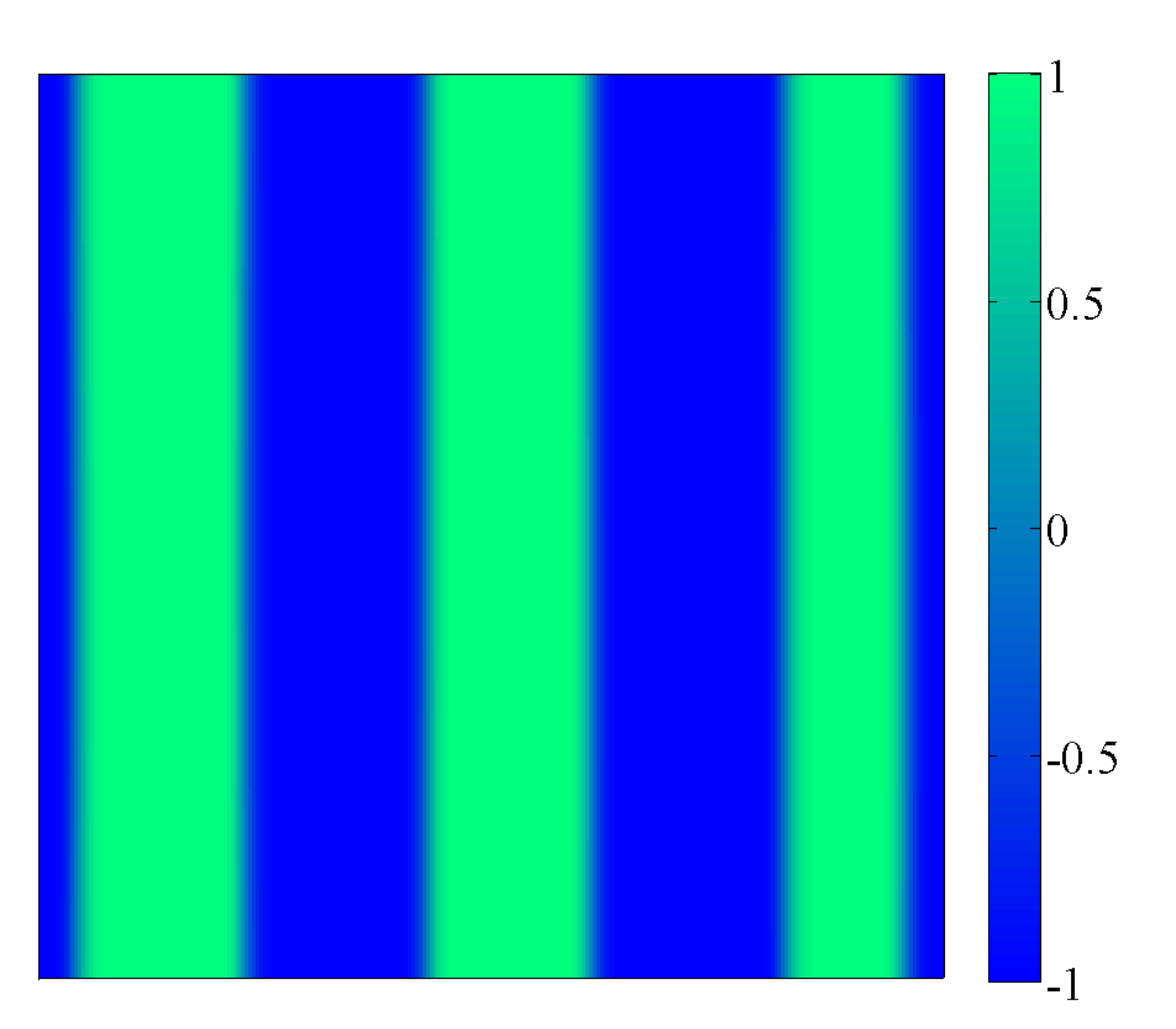}}
}
\caption{A snapshot of the steady-state composition for (a)
$V_0=10$; (b) $V_0=100$; (c) $V_0=1000$.}
\label{fig:C_const_flow}
\end{figure}
upper and lower bounds obtained in Eqs.~\eqref{eq:eqn_for_x}
and~\eqref{eq:funky_variance} are independent of the flow strength.
  
We solve Eq.~\eqref{eq:const_flow} numerically for various values of
$V_0$ and present the results in Fig.~\ref{fig:C_const_flow}.  For
small $V_0$, the composition morphology is similar to the flowless
case seen in Fig.~\ref{fig:no_flow}(d), except now the domains are
uniformly advected in a direction perpendicular to the source
variation.  The small-$V_0$ case is shown in
Fig.~\ref{fig:C_const_flow}(a).  As $V_0$ increases, the domain
boundaries are distorted due to the advection, while for large $V_0$,
the advection is sufficiently strong to break up the laminar domains.
The domain-like structure persists at late times however, and narrower
laminar domains form.

The $m_4$ measure of mixedness is almost constant across the range of
stirring parameters $0 \le V_0 \le 1000$.  For $V_0=1000$, $m_4$ is
slightly smaller than its value at $V_0=0$, due to the presence of
more interfaces.  This difference is small however, and increasing
$V_0$ does little to mix the fluid.  This is not surprising, given
that local shears are necessary to break up domain
structures~\cite{Berti2005}, and that such shears are absent from
constant flows.  What this example shows however, is the difference
between a miscible mixture, and a phase-separating mixture.  For a
diffusive mixture with the sinusoidal source we have studied, the
constant flow discussed here is optimal for
mixing~\cite{Thiffeault2007}; for a phase-separating mixture, the
constant flow badly fails to homogenize the mixture.

\section{Conclusions}

We have introduced the advective Cahn--Hilliard equation with a
mean-zero driving term as a way of describing a stirred, phase-separating
fluid, in the presence of sources and sinks.  By specializing to symmetric
mixtures, we have studied a more tractable problem, although one with many
applications.

Our goal was to investigate stirring protocols numerically and
analytically, and to determine the best way to break up the domains in
the Cahn--Hilliard fluid and achieve homogenization.  To this end, we
introduced the $m_p$ measure of composition fluctuations.  Since in a
well-mixed fluid, the composition exhibits spatial fluctuations about
the mean, with better mixing leading to smaller fluctuations, we used
$m_p$ as a measure of mixedness or homogeneity.  We proved the
existence of $m_p$ for long times, for $1 \le p \le 4$, and obtained
\emph{a priori} upper and lower bounds on $m_4$, as an explicit
function of the imposed flow $\bm{v}\left(\bm{x},t\right)$, and the
source $s\left(\bm{x}\right)$.
  
We compared the level homogeneity achieved by the random-phase sine
flow and the constant flow with the lower bound, and found that the
sine flow is effective at homogenizing the binary fluid, while the
constant flow fails in this task.  This is not surprising, since it is
known that differential shears are needed to break up binary fluid
domains, although it is radically different from the
advection-diffusion case, where the constant flow was the optimal
mixer.  The question of whether or not a flow is a good mixer in this
context was discussed using the mixing enhancement, defined in
Sec.~\ref{sec:numerics}.  Given such a definition, it is possible to
compare stirring protocols and find the optimal protocol for mixing a
binary fluid.  Our upper bound on the enhancement provides a meaningful
notion of this optimality.  This result may be useful in applications
where the homogenization of a binary fluid is desirable, since we have
set a lower limit on precisely how much homogeneity can be achieved.
  
L.\'O.N. was supported by the Irish government and the UK Engineering and
Physical Sciences Research Council. J.-L.T. was supported in part by the
UK EPSRC Grant No. GR/S72931/01.
 

\end{document}